\documentclass{article}

\usepackage[english]{babel}

\usepackage[letterpaper,top=2cm,bottom=2cm,left=3cm,right=3cm,marginparwidth=1.75cm]{geometry}

\usepackage{amsmath}
\usepackage{graphicx}
\usepackage[colorlinks=true, allcolors=blue]{hyperref}

\usepackage{lineno,bm,float,soul,amsfonts,amsmath,siunitx}
\usepackage{xcolor,hyperref}
\usepackage{empheq}
\usepackage{multirow}

\usepackage{adjustbox}
\usepackage{stmaryrd}

\usepackage{booktabs}
\usepackage{siunitx}

\newcommand{\T}{^{\mbox{\tiny T}}}

\newcommand{\B}[1]{{\bm #1}}

\title{Segmentation of the spacecraft transfer problem through overdetermined and continuity constraints based on the Theory of Functional Connections}
\author{A. K. de Almeida Jr.\footnote{CFisUC, Departamento de Física, Universidade de Coimbra, Portugal.\\
\href{https://orcid.org/0000-0002-9488-4462}{\textcolor{blue}{ORCID: 0000-0002-9488-4462}}\\
\href{https://scholar.google.com/citations?user=tgGQzGAAAAAJ}{\textcolor{blue}{Scholar: tgGQzGAAAAAJ}} }}
\begin{document}
\maketitle

\begin{abstract}
This paper introduces a segmented approach for solving constrained orbit transfer problems. The segments are connected through continuity constraints under the Theory of Functional Connections (TFC) mathematical framework that performs linear functional interpolation. This approach is further enhanced by a general vector formulation, from where the constrained functional is derived considering also a set of linear \textit{overdetermined constraints}. Since we constrain vector instead of coordinates, this methodology allows to apply TFC to multiple and complex constraints composed by any types of nonlinear components included. We demonstrate the effectiveness of the method on Earth-to-Moon transfers, showing that this segmented approach achieves solutions with several orders of magnitude greater accuracy and efficiency in comparison with unsegmented orbit transfers.
\end{abstract}

\section{Introduction}

The optimization of orbital transfers is a canonical problem in astrodynamics, typically formulated as a two-point boundary value problem (TPBVP). The design of minimum-cost Earth-to-Moon trajectories, in particular, has been a subject of extensive research, following early cost estimations by Sweetser \cite{sweetser1991estimate}. 
% Such a transfer can be designed with low thrust \cite{2018palau,doi:10.2514/2.4210,doi:10.2514/3.21515} or bi-impulsive Hohmann maneuver, solved via the patched restricted three-body problem \cite{da2012optimal,5586384}, full ephemeris $n$-body problem \cite{doi:10.2514/3.21079,LEI2013917}, restricted three body-problem \cite{pernicka95,YAGASAKI2004313,doi:10.2514/1.7702,da2012optimal}, restricted four-body problem \cite{ASSADIAN2010398}, and bi-circular restricted four-body problem 
% \cite{topputo2013optimal,Oshima2017,Onozaki2017,QI2017106,oshimatop19}.
Historically, shooting methods have been the predominant technique for solving such TPBVPs by iteratively solving a sequence of initial value problems. In this approach, the starting position is combined with a initial guess for the velocity and integrated for a given time of flight, and the final result is adopted to improve the initial velocity for the next iteration. Enhancements to this approach include gradient-based methods leveraging the Lambert problem for initial guess generation in perturbed environments \cite{pradobroucke}. Furthermore, low-energy Earth-Moon transfer is evaluated in \cite{gaggfilho2017} using the sequential gradient restoration algorithm \cite{miele1969}. To simplify this critical step, Gagg Filho and Fernandes \cite{gaggfilho2015, gaggfilho2016} developed patched-conic approximations based on the two-body problem to effectively provide an initial guess for more complex models of Earth-Moon missions. An extended version adopting variational equations of the Jacobi integral for Earth to Moon transfers is shown in \cite{GAGGFILHO2019312}. 
Modern procedures determine optimal impulsive trajectories by integrating the dynamics of the restricted three- or four-body problem with optimization techniques \cite{topputo2013a}. 
For instance, optimal transit orbits are determined within the four-body problem by dividing it into two three-body systems and employing methods such as Poincaré sections and the gradient method \cite{topputo2005b}. 
However, a significant and often difficult challenge in these approaches is establishing the initial guess for the optimization. 
The complexity of this task is highlighted in \cite{topputo2013optimal}, which required a four-dimensional space search just to generate an initial guess for Earth-Moon trajectories. This methodology was was based on the direct transcription and multiple shooting method, an improvement in segmenting the orbit based on a uniform time grid. 
This technique is later implemented in numerous subsequent studies involving transfers between Earth and Moon, e.g. \cite{oshimatop19}.
% This methodology is adopted in many subsequent studies, e.g. \cite{oshimatop19}. ,
% The choice of dynamical model is critical and spans a wide spectrum of complexities. Trajectories can be realized through low-thrust propulsion \cite{2018palau, doi:10.2514/2.4210, doi:10.2514/3.21515} or bi-impulsive Hohmann maneuvers, with the dynamics modeled under various assumptions. These include the patched restricted three-body problem \cite{da2012optimal, 5586384}, the high-fidelity full ephemeris n-body problem \cite{doi:10.2514/3.21079, LEI2013917}, the restricted three-body problem \cite{pernicka95, YAGASAKI2004313, doi:10.2514/1.7702, da2012optimal}, and different versions of the four-body problem \cite{ASSADIAN2010398, topputo2013optimal, Oshima2017, Onozaki2017, QI2017106, oshimatop19}.
In recent years, novel methodologies have been developed to solve this problem. Among these, the Theory of Functional Connections (TFC) has proven to be a particularly efficient and powerful framework in designing orbital transfers \cite{fastTFC}.

% \hline

The TFC is a mathematical framework \cite{U-ToC} to perform linear functional interpolation. Its core principle is the generation of analytical \textit{constrained functionals}, which are specifically designed to inherently satisfy a given set of linear constraints. By doing so, TFC transforms complex constrained problems into simpler, unconstrained ones, effectively reducing the solution search space.
Initially applied to linear differential equations and initial value problems (IVPs) using least-squares methods \cite{LDE}, the TFC framework was soon adapted for nonlinear problems as well \cite{NDE}. Its robustness and efficiency have been demonstrated in a wide array of fields, from solving differential equations \cite{Leake2019,florion2,Wang2024} and optimal control and trajectory optimization  \cite{Johnston2020,Li2021,Wang2022} to applications in particle physics \cite{Floriononame1,Schiassi2022}, biological modeling \cite{Xu2020,Daryakenari2024}, and geodesy \cite{Mortari2022}. The method's scope can be extended via coordinate transformations \cite{tfcvariables} and domain mapping \cite{bijective}. It has shown notable success in astrodynamics, tackling challenges like orbit determination \cite{akajtfcnelder}, Earth–Moon transfers \cite{fastTFC,akajtangential,Campana2024,tfc_connecting_earth_moon_2025}, and the computation of periodic orbits \cite{DEALMEIDA2023102068}.

In this work, we build upon the approach shown in \cite{fastTFC} to solve TPBVPs with TFC, extending its accuracy, robustness and applicability to complex boundary conditions. Our first contribution is the segmentation of the trajectory, where we divide the orbit into smaller pieces. These segments are then smoothly connected by imposing \textit{continuity constraints} on position and velocity at the junction between each pair of segments.
Our second contribution addresses a fundamental limitation: TFC is a framework to perform ``linear'' functional interpolation, whereas orbit transfer boundaries more complex than TPBVPs are often nonlinear. To overcome this without resorting to limited and potentially complicated coordinate transformations \cite{tfcvariables} to linearize the constraints, we introduce a novel set of \textit{overdetermined constraints}. This set of constraints allows to use TFC imposing multiple and complex constraints. For instance, we impose a pair of radius denoting circular orbits and a pair of tangential velocities constraints at both the initial and final time of motion, which was not possible using transformation to polar coordinates centered at Moon as shown in \cite{akajtangential} - only one tangential velocity constraint is possible using this transformation. The \textit{overdetermined constraints} are obtained using a vector formulation for TFC, thus extending the formulation shown in \cite{tfc_solarsail}. This novel approach is very important, because both these \textit{overdetermined} and the \textit{continuity} constraints are linear within our proposed vector formulation. Hence, the \textit{constrained functionals} can be derived using TFC, as we show here.

% In this paper, we greatly extend the approach shown \cite{fastTFC} to solve the TPBVP using TFC by proposing segmentation of the orbit to increase accuracy, while also extending it to more general and complex boundaries. After segmenting the orbit, we show how to connect these segments through a set of \textit{continuity constraints}. These constraints smoothly connect the segments, imposing continuities on positions and velocities at the connecting time of every pair of segments. Furthermore, we propose in this paper a set of \textit{overdetermined constraints}. As explained above, TFC is a framework to perform linear functional interpolation. On the other side, in a boundary value problem, the constraints associated to orbit transfers are often nonlinear. TFC in principle could not be applied to such a class of problems. Although a transformation of coordinates is possible to contour this problem \cite{tfcvariables}, solving the differential equations in some system of coordinates may be numerically problematic. In this paper, we present a novel approach. We take advantage of TFC to derive the constrained functional for the \textit{overdetermined constraints} combined with \textit{continuity constraints} based on a vector formulation, extending then the formulation firstly shown in \cite{tfc_solarsail}. This is possible because both the \textit{continuity} and \textit{overdetermined} constraints are linear in this formulation, and TFC can indeed handle linear constraints. 

Although we base applications to astrodynamics (in segmentation of orbits), the methodology is presented in a general vector formulation, completely independent of the equations of the dynamics. This means that the technique presented here can be applied to a broad range of systems subject to boundary conditions from areas such as biology, physics, chemistry, engineering, etc. We illustrate application and investigate numerical comparisons between segmented and non-segmented orbits, analyzing the gains related to different numbers of segmentation. Such an analysis shows a significant gain in accuracy in segmenting the orbit with respect to unsegmented traditional orbit transfer, e.g. as those obtained in \cite{fastTFC}. The proposed \textit{overdetermined constraints} drastically facilitates numerical convergence in comparison with component constraints, e.g. those associated to the TPBVP as adopted in \cite{fastTFC}. Thus, solutions are more easily obtained for longer times of flight. Analyses showed that accuracy increases with more segmented Earth to Moon orbit transfer. Since segmenting orbits also increases the computational load, we define a relative computational cost to estimate the computational gain in the process. 
% Although this index does not consider the numerical load due to the complexity of the constrained functional, it 
This index takes into consideration the numerical computational load due to the increasing in the size of the Jacobian matrix, whose inverse (or pseudo-inverse) is evaluated in the optimization process. The relative computational cost shows a significant gain of several orders of magnitude in segmenting the orbit. Thus, segmenting orbits is recommended even in the case where this increasing in the computational load is considered.

\section{Mathematical formulation}

%The time range may be numerically limited. 
%Machine accuracy level  
%we propose to extend the time range accuracy by dividing the time of flight into segments. The solution is then obtained for the complete system.
%We start the procedure dividing the time of flight into smaller segments, and they will be connected later through continuity constraints.

\begin{figure}
\centering\includegraphics[scale=0.45]{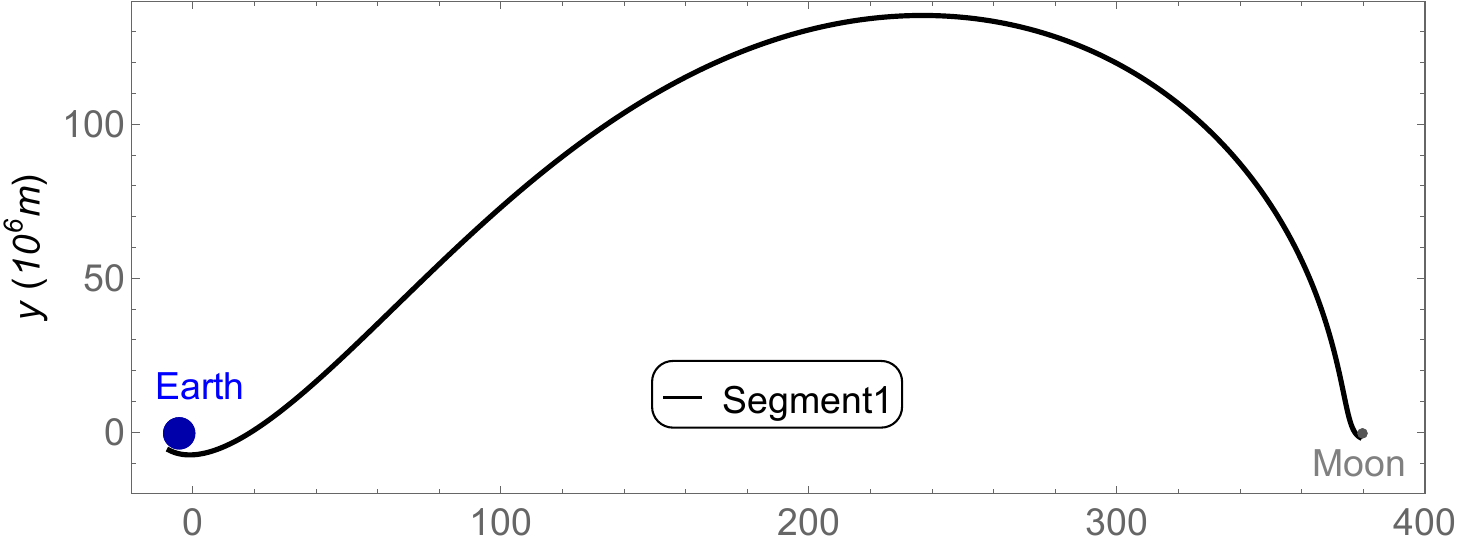}
\centering\includegraphics[scale=0.45]{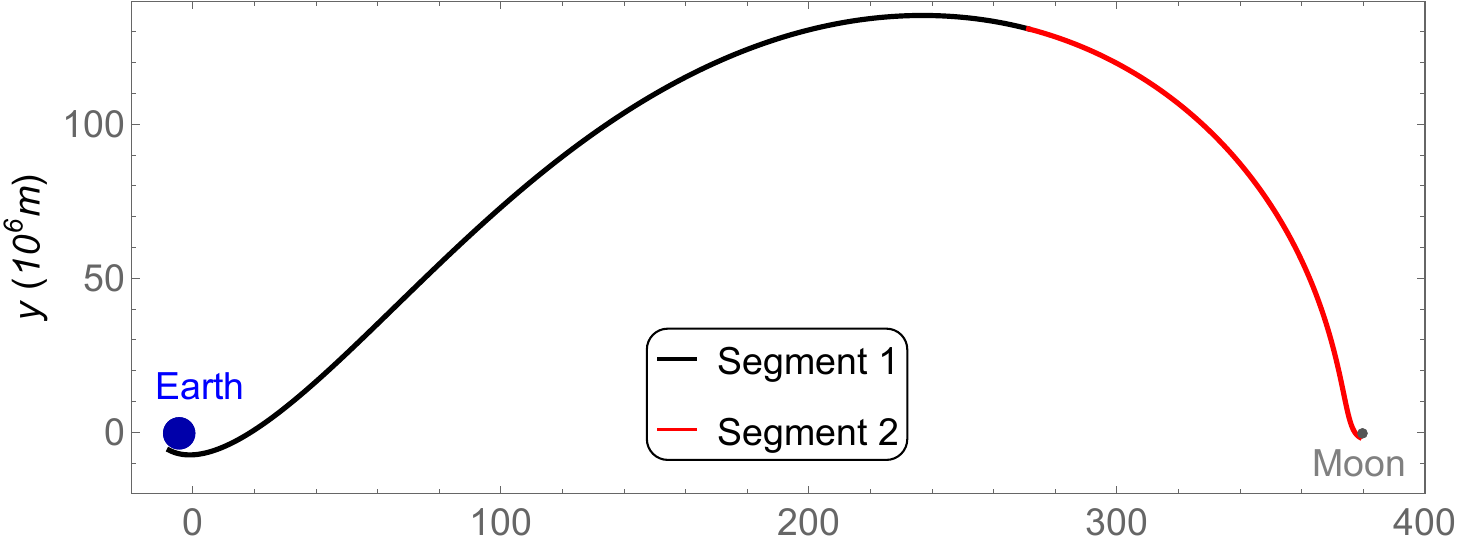}
\centering\includegraphics[scale=0.45]{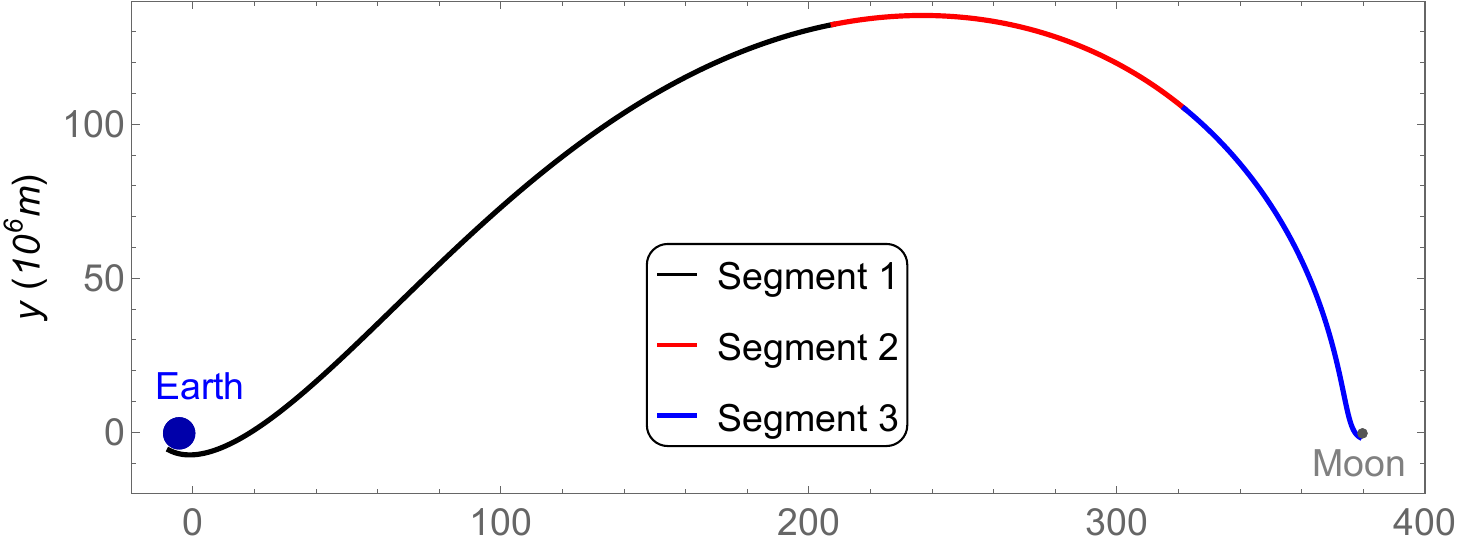}
\centering\includegraphics[scale=0.45]{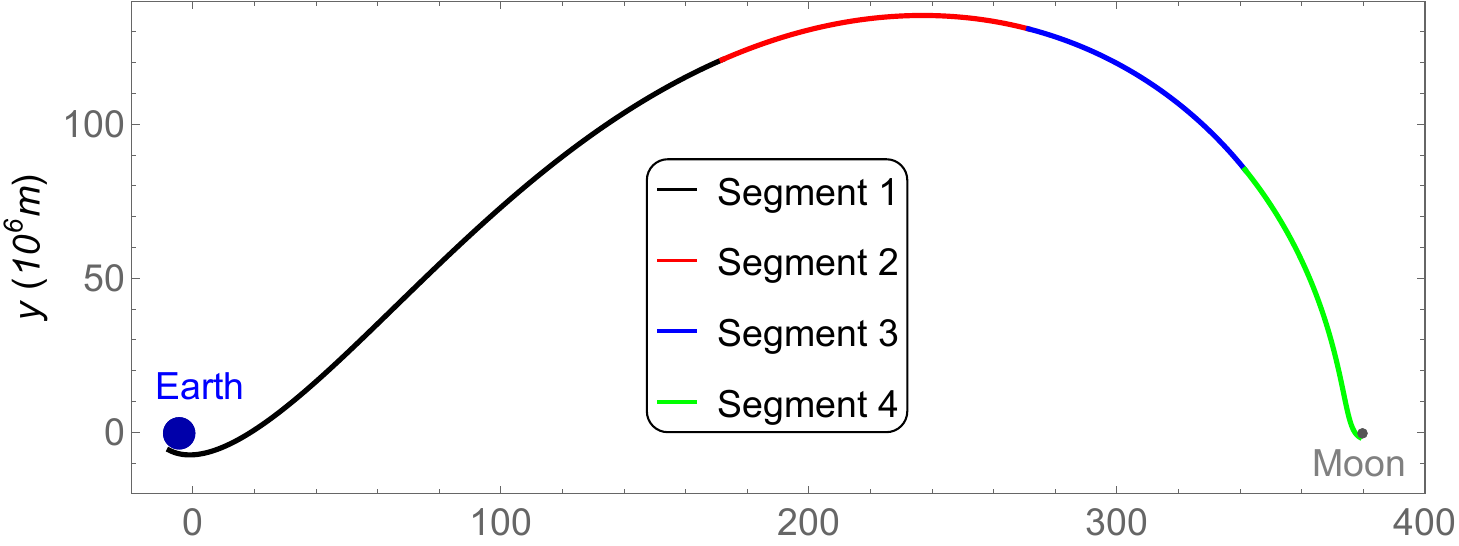}
\centering\includegraphics[scale=0.45]{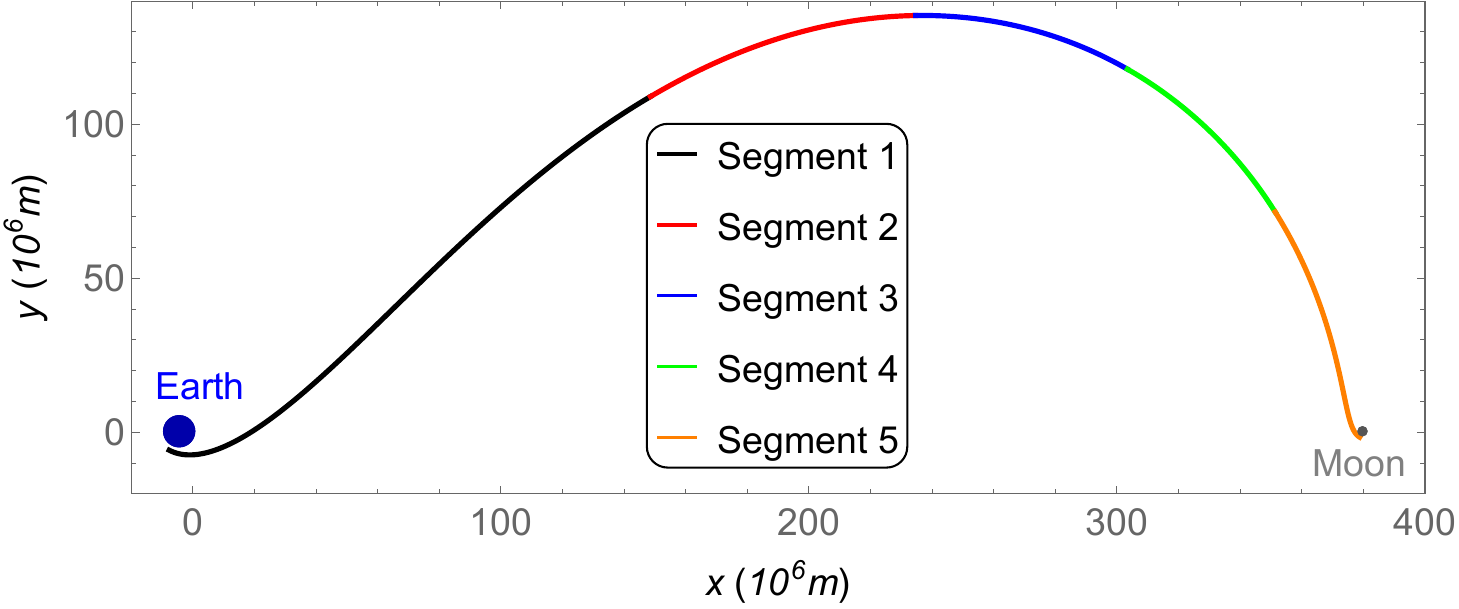}
\caption{Earth to Moon orbit transfer using different numbers of segmentation up to 5.}
\label{fig:Segments}
\end{figure}

We start the procedure dividing the time of flight into smaller segments. 
The time $\tau$ varies from $\tau_i$ to $\tau_f$, where the total time of flight is then $T_f=\tau_f-\tau_i$. We divide the trajectory into $N_s$ segments equally spaced in time, as illustrated in Fig. \ref{fig:Segments} to the case of a transfer from Earth to Moon for several values of $N_s$ in the interval $\llbracket 1:5\rrbracket$. It can be noted that the first case for $N_s=1$ has only one segment.
%, i.e. it is equivalent to traditional not segmented solution.
Hence, the time of flight of every segment is 
\begin{equation}\label{eq:T}
    T=\frac{T_f}{N_s}
\end{equation}
and the time $\tau$ of the $n^{th}$ segment varies from $\tau=(n-1)T_f/N_s$ to $\tau=n T_f/N_s$.
On the other hand, without loss of generality, we adopt the time $t_n$ for each of these trajectories through the following time transformation for the $n^{th}$ segment:
\begin{equation}\label{eq:ttau}
 t_n = \tau - \tau_i - \frac{(n-1)T_f}{N_s}
\end{equation}
This is an important step, and we adopt this formulation because the solution of all segments must be obtained simultaneously. Note from the time transformation shown in Eq. \eqref{eq:ttau} that $t_n$ varies from $0$ to $T$. Thus, when we vary the time $t_n$ from $0$ to $T$ for every one of the $N_s$ segments simultaneously, we are obtaining the solution for every segment, and, hence, for the complete motion by composing them back according to Eq. \eqref{eq:ttau}. This composition is clearly shown in Fig.~\ref{fig:Segments}. Since these parts of time $t_n$ are equivalent, i.e. they vary from $0$ to $T$ under the same scale, from now on, without loss of generality, we adopt the notation $t$, instead of $t_n$, for the time of every segment for simplification purposes.

The position $\B{r}$ is decomposed as a matrix of order $3 \times N_s$, where the lines denotes its components (depending on the system of coordinates), and the columns its segmentation into $N_s$ parts. The composition of the position $\B{r}$ is then shown as
%is then composed with the segmentation described above according to
\begin{equation}\label{eq:r}
    \B{r}= \, \begin{Bmatrix} \B{r}_1, &...&, \B{r}_{n-1},& \B{r}_{n},& \B{r}_{n+1},&...&, \B{r}_{N_s} \end{Bmatrix} ,
    % \B{r}= \, \begin{Bmatrix} \B{r}_1, &\B{r}_2, &...&, \B{r}_{n-1},& \B{r}_{n},& \B{r}_{n+1},&...&, \B{r}_{N_s-1}, & \B{r}_{N_s} \end{Bmatrix}\T 
    %= \begin{Bmatrix} x_1, & y_1, & z_1, & x_2, &  y_2, & z_2\end{Bmatrix}\T
\end{equation}
where $\B{r}_1$ is a three dimensional vector describing the motion in the first segment, $\B{r}_2$ describes the motion in the second segment, and so on, up to $\B{r}_{N_S}$, which describes the motion for the last segment. 
% The time interval is the same for all segments, varying from $t_n=0$ to $t_n=T$ for the $n^{th}$ segment. Thus, without loss of generality and for simplicity of notation, we eliminate the index notation $n$ from the time, and call it $t$ for every segment. Note that $t$ varies then from $0$ to $T$.
%Thus, if we want to compose the complete motion using these segments, we must consider that the time of the second segment starts at the final time of the first segment, and so on.

% \subsection{The TFC formulation}

It can be noted from the notation defined above that every segment $\B{r}_n$ for $n\in\llbracket 1:N_s\rrbracket$ is a three dimensional vector written as a matrix of order $3\times1$. The constrained functional can be obtained from the $\eta$ formulation of TFC \cite{U-TFC} for the three-dimensional $n^{th}$ segment vector from the expression \cite{tfc_solarsail}
% \begin{equation}\label{eq:generating2}
%     \B{r} = \B{g} + E \, \B{s},
% \end{equation}
\begin{equation}\label{eq:generating_i0}
    \B{r}_n = \B{g}_n + E_n \, \B{s}_n, \quad \text{for} ~ n\in\llbracket 1:N_s \rrbracket
\end{equation}
where $\B{g}_n = \B{g}_n (t)$ is a column vector of free functions of order $3 \times 1$, $E_n$ is a constant matrix of order $3 \times k_n$, and $\B{s} = \B{s} (t)$ is a vector of support functions of order $k_n \times 1$, where $k_n$ is the number of support functions adopted for the $n^{th}$ segment. Due to the reason that we integrate the system (we connect the segments), in this paper, the number of support functions $k_n$ may be different from the number of constraints related to its segment, although the total number of support functions for the integrated system is equal to the number of constraints of the integrated system, called as ``combined constraints''. This generalization will be explained in Sec. \ref{sec:segNs2}.

\subsection{Connecting the segments}
\label{sec:application}

In order to integrate the problem, trajectory segment $\B{r}_n$ and trajectory segment $\B{r}_{n+1}$ must be unified at a specific point. The segment $\B{r}_n$ is connected to the next segment $\B{r}_{n+1}$ through the following constraints:
\begin{equation}\label{eq:continuity}
\text{Continuity constraints:}
\begin{cases}
    \B{r}_n(T)=\B{r}_{n+1}(0) \\
    \dot{\B{r}}_n(T)=\dot{\B{r}}_{n+1}(0),
\end{cases}
\end{equation}
for $n\in\llbracket 1:N_s-1 \rrbracket$. These constraints guarantee smooth continuity between the solutions obtained for trajectories $\B{r}_n$ and $\B{r}_{n+1}$. 
In the next sections, we take advantage of the TFC to integrate these continuity constraints into the transfer problem.

% \subsection{Case \texorpdfstring{$N_s=2$}{Ns=2}}

% \subsubsection{Nonlinear components constraints}
\subsection{The overdetermined constraints}
\label{sec:oc}

\begin{figure}[t]
    \centering
    \includegraphics[scale=0.4]{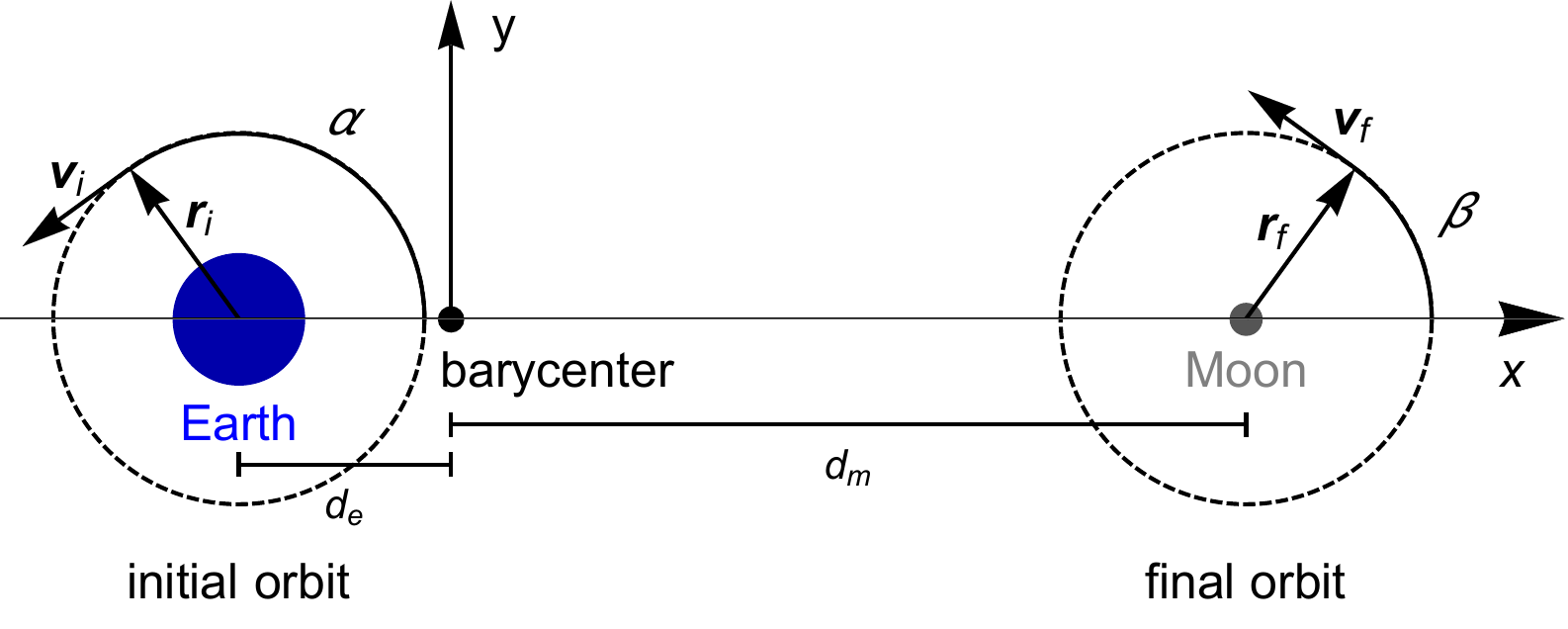}
    \caption{An example of a rectangular system of coordinates ($x,y$) centered at the barycenter, where Earth and Moon are located along the $x$ axis.}
    \label{fig:graph1}
\end{figure}

Although the constraints depend on the problem to be solved, we illustrate the methodology by applying it to design transfers between Earth and Moon. Thus, we show how to adopt \textit{overdetermined constraints} to this purpose.
%We illustrate them by applying it to design transfers between Earth and Moon.
% in the particular case where $N_s=2$, and then we extend it to larger values of $N_s$ in subsequent sections. 
% In this application, 
In such a transfer, the spacecraft is initially in a circular orbit of radius $r_e$ around the Earth and is transferred to another circular final orbit around the Moon with radius $r_m$ in a time of flight $T_f$. This transfer is done through a bi-impulsive maneuver. An optimum solution coming from the 2-body problem approximation is applying the impulses tangentially to the circular orbits. This transfer can be performed in the plane of motion of the pair Earth-Moon. In rectangular coordinates, where $\B{r}_1=\{x_1,y_1\}$ and $\B{r}_2=\{x_2,y_2\}$, these four constraints are written as
\renewcommand{\theenumi}{\roman{enumi}}
\begin{enumerate}
	\item $\sqrt{(x_1+d_e)^2 + y_1^2}|_{t = 0} = r_e$
	\item $\sqrt{(x_2-d_m)^2 + y_2^2}|_{t = T} = r_m$
	\item $(\dot{x}_1 \, (x_1+d_e) + \dot{y}_1 \, y_1)|_{t = 0} = 0$
    \item $(\dot{x}_2 \, (x_2-d_m) + \dot{y}_2 \, y_2)|_{t = T} = 0$
\end{enumerate}
where $d_e$ and $d_m$ are the distance of Earth and Moon to their common barycenter located at the center of this system of coordinates, as depicted in Fig. \ref{fig:graph1}.
The problem is that these constraints are not linear in the variables, while TFC framework has been fully developed only for linear constraints. Thus, TFC could not be applied to these types of constraints. This issue can be addressed via a change of variables \cite{tfcvariables}, e.g. using polar coordinates centered at Moon \cite{akajtangential}, although in this last case, only the final velocity is tangential to the orbit. The initial tangential velocity cannot be linearly constrained using polar coordinates centered at the Moon. In general, only few constraints can become linear with a change of coordinates, and the symmetry is often not clear to decide what type of change of variables should be used to this task. In the next paragraph, we show how to address this problem with no need of change of variables. In fact, the novel formulation presented in this paper allows to embed both the initial and final velocities to be tangentially constrained, additionally to constrain both altitudes of the orbit, around Earth at starting time and Moon at final time. These aspects will be better explained later in Sec. \ref{sec:tangential}.

% \subsubsection{Overdetermined constraints}
% \label{sec:oc}

We take advantage of the general vector formalism presented in this paper, where we constrain vectors instead of coordinates. This general formulation is better because vectors can be written in any types of coordinates, e.g. using rectangular coordinates. We adopt this novel approach using overdetermined constraints for the initial and final position ($\B{r}_i$ and $\B{r}_f$, respectively) and velocities ($\B{v}_i$ and $\B{v}_f$, respectively). We call them overdetermined constraints because the initial (or final) position and velocity should be sufficient to determine problem, whereas here we add two more constraints. The overdetermined constraints are then defined as
\begin{equation}\label{eq:over_constr}
\text{Overdetermined constraints:}
\begin{cases}
    \B{r}_1(0)=\B{r}_i \\
    \dot{\B{r}}_1(0)=\B{v}_i\\
    \B{r}_{N_s}(T)=\B{r}_f \\
    \dot{\B{r}}_{N_s}(T)=\B{v}_f.
\end{cases}
\end{equation}
Although these constraints seems superposed, the vectors $\B{r}_i$, $\B{r}_f$, $\B{v}_i$, and $\B{v}_f$ will be just partially (and strategically) defined. 
Beyond being linear, these constraints are capable of containing embedded not only unknowns variables, 
%We take advantage of this capability to embed on them not only unknowns, 
but also multiple and complex nonlinear constraints in the vector components, as will be explored later in Sec. \ref{sec:tangential}. Moreover, since they are linear, the constrained functional can be obtained using TFC, as shown next.

\subsection{Unsegmented case}

% We illustrate our method by applying it to design transfers between Earth and Moon in the particular case where $N_s=2$, and then we extend it to larger values of $N_s$ in subsequent sections. 

This is the most simple case shown in this paper representing a not segmented orbit transfer designed under the \textit{overdetermined constraints} shown in Eq.~\eqref{eq:over_constr} for $N_s=1$. We use the support function given by $\B{s}_1=\begin{Bmatrix}1,t,t^2,t^3\end{Bmatrix}\T$ 
%to derive the 
%We use the support functions shown above (in Eq.~\eqref{eq:sf3}) 
into Eq.~\eqref{eq:generating_i0}, and then apply Eq.~\eqref{eq:generating_i0} into Eq.~\eqref{eq:over_constr}. The \textit{constrained functional} is obtained by solving this linear system for $E_1$ and replacing it into Eq.~\eqref{eq:generating_i0} as

\begin{eqnarray}\label{eq:Ns1} \nonumber
        	\B{r}_1(t)&=&\B{g}_1(t)-\B{g}_1(0)+\B{r}_i+t \left(\B{v}_i-\dot{\B{g}}_1(0)\right)\\ 
            &&+t^2 \left(\frac{\dot{\B{g}}_1(T)+2 \dot{\B{g}}_1(0)-2 \B{v}_i-\B{v}_f}{T}-\frac{3 (\B{g}_1(T)-\B{g}_1(0)+\B{r}_i-\B{r}_f)}{T^2}\right)\\ \nonumber
            &&+t^3 \left(\frac{-\dot{\B{g}}_1(T)-\dot{\B{g}}_1(0)+\B{v}_i+\B{v}_f}{T^2}+\frac{2 (\B{g}_1(T)-\B{g}_1(0)+\B{r}_i-\B{r}_f)}{T^3}\right).
\end{eqnarray}

\subsection{Segmenting into two parts}
\label{sec:segNs2}

% In the results, we show comparison with traditional cases, e.g. \cite{fastTFC}, where the boundary value problem is not segmented, i.e. $N_s=1$.
In the case where $N_s=2$, the segment is divided in two by halving the total time of flight. Thus, the partial time of flight of trajectory 1 (between starting and half times) is equal to the time of flight of trajectory 2 (between half and final times). The time for each of these trajectories is from $0$ to $T=T_f/2$, according to Eq. \eqref{eq:T} for $N_s=2$. We connect trajectory 1 with trajectory 2 using the continuity constraints shown in \eqref{eq:continuity} for $n=1$, guaranteeing smooth continuity between trajectories 1 and 2. This connection can be visualized in Fig. \ref{fig:Segments}, where the second plot (from up to down) represents the $N_s=2$ case.

% In order to integrate the problem, trajectory 1 and trajectory 2 must be unified. In this paper, 
% We connect trajectory 1 with trajectory 2 using the following constraints:
% \begin{equation}\label{eq:continuityN_s2}
% \begin{cases}
%     \B{r}_1(T)=\B{r}_2(0) \\
%     \dot{\B{r}}_1(T)=\dot{\B{r}}_2(0).
% \end{cases}
% \end{equation}
% These constraints guarantee smooth continuity between the solutions obtained for trajectories 1 and 2. 

% \section{Overdetermined constraints}

% \subsection{Segmenting and connecting the orbit transfer problem}
% \label{sec:application}

Combining the overdetermined constraints shown in Eq.~\eqref{eq:over_constr} with the continuation constraints shown in \eqref{eq:continuity} for $n=1$, each segment is then subject to the following four constraints:
\begin{equation}\label{eq:constr_N2}
\text{Segment 1:}
\begin{cases}
    \B{r}_1(0)=\B{r}_i \\
    \dot{\B{r}}_1(0)=\B{v}_i\\
    \B{r}_1(T)=\B{r}_2(0) \\
    \dot{\B{r}}_1(T)=\dot{\B{r}}_2(0)
\end{cases}
\text{Segment 2:}
\begin{cases}
    \B{r}_2(T)=\B{r}_f \\
    \dot{\B{r}}_2(T)=\B{v}_f\\
    \B{r}_1(T)=\B{r}_2(0) \\
    \dot{\B{r}}_1(T)=\dot{\B{r}}_2(0)
\end{cases}
\end{equation}
Although each segment is subject to 4 constraints, the two continuity constraints are shared between the two segments. Our methodology combines the segments into a complete system. The combined system is then subject to the 6 following constraints:
\begin{equation}\label{eq:constr_N2}
\text{Combined constraints:}
\begin{cases}
    \B{r}_1(0)=\B{r}_i \quad \quad \quad \quad \quad \quad
    \dot{\B{r}}_1(0)=\B{v}_i\\
    \B{r}_1(T)=\B{r}_2(0) \quad \quad \quad \quad \quad
    \dot{\B{r}}_1(T)=\dot{\B{r}}_2(0)\\
    \B{r}_2(T)=\B{r}_f \quad \quad \quad \quad \quad \quad
    \dot{\B{r}}_2(T)=\B{v}_f
\end{cases}
\end{equation}
Each segment is written according to \eqref{eq:generating_i0}, where in this work we adopt $\B{s}_n=\begin{Bmatrix}1,t,t^2\end{Bmatrix}\T$ for $n\in\llbracket 1:2 \rrbracket$.
It is important to note that, since we have 6 constraints for the complete system, the sum of the number of support functions adopted for the two segments must be equal to 6, i.e. $k_1+k_2=6$. Furthermore, one can choose different numbers of support functions for each segment. In fact, this number of support functions can be arbitrarily shared among the segments. For instance, we could have adopted the combination $k_1=4$ and $k_2=2$ representing $\B{s}_1=\begin{Bmatrix}1,t,t^2,t^3\end{Bmatrix}\T$ and $\B{s}_2=\begin{Bmatrix}1,t\end{Bmatrix}\T$, instead of $\B{s}_1=\B{s}_2=\begin{Bmatrix}1,t,t^2\end{Bmatrix}\T$ for $k_1=k_2=3$.
%Due to this reason, we use a number of support functions not equal to the number of constraints, but equal to the number of constraints minus the number of shared constraints divided by 2. 
%In case this number is not an integer, it is not a problem, because one can choose different numbers of support functions for each segment. 
%Although every segment shown above is subject to 4 constraints in a total of 12 combined constraints, four of these 12 constraints are repeated. Thus, the integrated system is subject to the following 8 constraints:
%Due to the reason we solve the integrated system, we should then use and distribute 8 support functions among the segments. In this paper, we adopt the following distribution
The constrained functionals are obtained combining Eqs.~\eqref{eq:generating_i0} and \eqref{eq:constr_N2}, solving for $E_1$ and $E_2$ simultaneously, and replacing them into Eq. \eqref{eq:generating_i0}. The result is
\begin{equation}\label{eq:cfN2}
    \begin{cases}
        \B{r}_1=\B{g}_1(t)-\B{g}_1(0)+\B{r}_i+ \left(\B{v}_i-\dot{\B{g}}_1(0)\right) t\\
        -\left(-3 T \dot{\B{g}}_1(0)+T \dot{\B{g}}_1(T)+2 \B{g}_1(T)-2 \B{g}_1(0)-T \dot{\B{g}}_2(0)-T \dot{\B{g}}_2(T)+2 \B{g}_2(T)-2 \B{g}_2(0)+3 T \B{v}_i+T \B{v}_f+2 \B{r}_i-2 \B{r}_f\right)(\frac{t}{2T})^2\\
        \B{r}_2=\B{g}_2(t)\\
        +\left(-T \dot{\B{g}}_1(0)-T \dot{\B{g}}_1(T)+2 \B{g}_1(T)-2 \B{g}_1(0)+T \dot{\B{g}}_2(0)+T \dot{\B{g}}_2(T)-2 \B{g}_2(T)-2 \B{g}_2(0)+T \B{v}_i-T \B{v}_f+2 \B{r}_i+2 \B{r}_f\right)\frac{1}{4}\\
        -\left(-T \dot{\B{g}}_1(0)-T \dot{\B{g}}_1(T)+2 \B{g}_1(T)-2 \B{g}_1(0)+T \dot{\B{g}}_2(0)-T \dot{\B{g}}_2(T)+2 \B{g}_2(T)-2 \B{g}_2(0)+T \B{v}_i+T \B{v}_f+2 \B{r}_i-2 \B{r}_f\right) \frac{t}{2T}\\
        -\left(T \dot{\B{g}}_1(0)+T \dot{\B{g}}_1(T)-2 \B{g}_1(T)+2 \B{g}_1(0)-T \dot{\B{g}}_2(0)+3 T \dot{\B{g}}_2(T)-2 \B{g}_2(T)+2 \B{g}_2(0)-T \B{v}_i-3 T \B{v}_f-2 \B{r}_i+2 \B{r}_f\right)(\frac{t}{2T})^2     
    \end{cases}
\end{equation}
The segmented positions $\B{r}_1$ and $\B{r}_2$ are then combined to form $\B{r}$, which has six dependent position coordinates (three for $\B{r}_1$ and three for $\B{r}_2$), according to Eq.~\eqref{eq:r}.

% \subsection{Segmentation and connection for \texorpdfstring{$N_s=3$}{Ns=3}}
\subsection{Segmenting into three parts}
\label{sec:segNs3}

In this particular case we divide the orbit into three segments. They are subject to the continuity constraints shown in Eq.~\eqref{eq:continuity} and to the overdetermined constraints shown in Eq.~\eqref{eq:over_constr}. Thus, the segments are subject to the following constraints:
\begin{equation}\label{eq:constr_N3a}
\text{Segment 1:}
\begin{cases}
    \B{r}_1(0)=\B{r}_i \\
    \dot{\B{r}}_1(0)=\B{v}_i\\
    \B{r}_1(T)=\B{r}_2(0) \\
    \dot{\B{r}}_1(T)=\dot{\B{r}}_2(0)
\end{cases}
\text{Segment 2:}
\begin{cases}
    \B{r}_1(T)=\B{r}_2(0) \\
    \B{r}_2(T)=\B{r}_3(0) \\
    \dot{\B{r}}_1(T)=\dot{\B{r}}_2(0)\\
    \dot{\B{r}}_2(T)=\dot{\B{r}}_3(0)
\end{cases}
\text{Segment 3:}
\begin{cases}
    \B{r}_3(T)=\B{r}_f \\
    \dot{\B{r}}_3(T)=\B{v}_f\\
    \B{r}_2(T)=\B{r}_3(0) \\
    \dot{\B{r}}_2(T)=\dot{\B{r}}_3(0).
\end{cases}
\end{equation}
Although every segment shown above is subject to 4 constraints in a total of 12 combined constraints, four of these 12 constraints are repeated. Thus, the integrated system is subject to the following 8 constraints:
\begin{equation}\label{eq:constr_N3b}
\text{Combined constraints:}
\begin{cases}
    \B{r}_1(0)=\B{r}_i \quad \quad \quad \quad \quad \quad
    \dot{\B{r}}_1(0)=\B{v}_i\\
    \B{r}_1(T)=\B{r}_2(0) \quad \quad \quad 
    \dot{\B{r}}_1(T)=\dot{\B{r}}_2(0)\\
    \B{r}_2(T)=\B{r}_3(0) \quad \quad \quad 
    \dot{\B{r}}_2(T)=\dot{\B{r}}_3(0)\\
    \B{r}_3(T)=\B{r}_f \quad \quad \quad \quad \quad 
    \dot{\B{r}}_3(T)=\B{v}_f
\end{cases}
\end{equation}
Due to the reason we solve the integrated system, we should then use and distribute 8 polynomials among the three support functions associated to the three segments. It should be noted that some combinations may lead to a non-solvable linear system in $E_n$. In this paper, we adopt the following distribution:
\begin{equation}\label{eq:sf3}
    \B{s}_1=\begin{Bmatrix}1,t,t^2\end{Bmatrix}\T; \quad 
    \B{s}_2=\begin{Bmatrix}1,t\end{Bmatrix}\T; \quad \text{and} \quad
    \B{s}_3=\begin{Bmatrix}1,t,t^2\end{Bmatrix}\T.
% \text{Support functions:}
% \begin{cases}
%     \B{s}_1=\begin{Bmatrix}1,t,t^2\end{Bmatrix}\T\\
%     \B{s}_2=\begin{Bmatrix}1,t\end{Bmatrix}\T\\
%     \B{s}_3=\begin{Bmatrix}1,t,t^2\end{Bmatrix}\T.
% \end{cases}
\end{equation}
We use the support functions shown above (in Eq.~\eqref{eq:sf3}) into Eq.~\eqref{eq:generating_i0}, and then apply Eq.~\eqref{eq:generating_i0} into \eqref{eq:constr_N3b} to generate a linear system in $E_n$ for $n\in\llbracket 1:3 \rrbracket$. The \textit{constrained functional} is obtained by solving this linear system for $E_n$ and substituting the solution into Eq.~\eqref{eq:generating_i0} as
\begin{equation}\label{eq:cfN3}
    \begin{cases}
        \B{r}_1=t \left(\B{v}_i-\dot{\B{g}}_1(0)\right)+\B{g}_1(t)-\B{g}_1(0)+\B{r}_i\\
        +t^2 \left(\frac{-5 \dot{\B{g}}_1(T)+8 \dot{\B{g}}_1(0)-2 \dot{\B{g}}_2(T)+5 \dot{\B{g}}_2(0)+\dot{\B{g}}_3(T)+2 \dot{\B{g}}_3(0)-8 \B{v}_i-\B{v}_f}{13 T}-\frac{3 (\B{g}_1(T)-\B{g}_1(0)+\B{g}_2(T)-\B{g}_2(0)+\B{g}_3(T)-\B{g}_3(0)+\B{r}_i-\B{r}_f)}{13 T^2}\right)\\
        
        \B{r}_2=\frac{1}{13} T \left(-5 \dot{\B{g}}_1(T)-5 \dot{\B{g}}_1(0)-2 \dot{\B{g}}_2(T)+5 \dot{\B{g}}_2(0)+\dot{\B{g}}_3(T)+2 \dot{\B{g}}_3(0)+5 \B{v}_i-\B{v}_f\right)\\
        +\frac{1}{13} (10 \B{g}_1(T)-10 \B{g}_1(0)+13 \B{g}_2(t)-3 \B{g}_2(T)-10 \B{g}_2(0)-3 \B{g}_3(T)+3 \B{g}_3(0)+10 \B{r}_i+3 \B{r}_f)\\
        + t \Big[\frac{1}{13} \left(3 \dot{\B{g}}_1(T)+3 \dot{\B{g}}_1(0)-4 \dot{\B{g}}_2(T)-3 \dot{\B{g}}_2(0)+2 \dot{\B{g}}_3(T)+4 \dot{\B{g}}_3(0)-3 \B{v}_i-2 \B{v}_f\right)\\
        -\frac{6 (\B{g}_1(T)-\B{g}_1(0)+\B{g}_2(T)-\B{g}_2(0)+\B{g}_3(T)-\B{g}_3(0)+\B{r}_i-\B{r}_f)}{13 T}\Big]\\
        
        \B{r}_3=\frac{1}{13} T \left(-2 \dot{\B{g}}_1(T)-2 \dot{\B{g}}_1(0)-6 \dot{\B{g}}_2(T)+2 \dot{\B{g}}_2(0)+3 \dot{\B{g}}_3(T)+6 \dot{\B{g}}_3(0)+2 \B{v}_i-3 \B{v}_f\right)\\
        +\frac{1}{13} (4 \B{g}_1(T)-4 \B{g}_1(0)+4 \B{g}_2(T)-4 \B{g}_2(0)+13 \B{g}_3(t)-9 \B{g}_3(T)-4 \B{g}_3(0)+4 \B{r}_i+9 \B{r}_f)\\
        +t \Big[\frac{1}{13} \left(3 \dot{\B{g}}_1(T)+3 \dot{\B{g}}_1(0)+9 \dot{\B{g}}_2(T)-3 \dot{\B{g}}_2(0)+2 \dot{\B{g}}_3(T)-9 \dot{\B{g}}_3(0)-3 \B{v}_i-2 \B{v}_f\right)\\
        -\frac{6 (\B{g}_1(T)-\B{g}_1(0)+\B{g}_2(T)-\B{g}_2(0)+\B{g}_3(T)-\B{g}_3(0)+\B{r}_i-\B{r}_f)}{13 T}\Big]\\
        +t^3 \left(\frac{-\dot{\B{g}}_1(T)-\dot{\B{g}}_1(0)-3 \dot{\B{g}}_2(T)+\dot{\B{g}}_2(0)-5 \dot{\B{g}}_3(T)+3 \dot{\B{g}}_3(0)+\B{v}_i+5 \B{v}_f}{13 T^2}+\frac{2 (\B{g}_1(T)-\B{g}_1(0)+\B{g}_2(T)-\B{g}_2(0)+\B{g}_3(T)-\B{g}_3(0)+\B{r}_i-\B{r}_f)}{13 T^3}\right)
    \end{cases}
\end{equation}

% \subsection{Segmentation and connection for \texorpdfstring{$N_s=4$}{Ns=4}}
\subsection{Segmenting into four parts}
\label{sec:segNs4}

In the case the orbit is divided into four segments, then $N_s=4$ and every segment is subject to the continuity constraints shown in Eq.~\eqref{eq:continuity} and to the overdetermined constraints shown in Eq.~\eqref{eq:over_constr}. The segments are then subject to the following four constraints:
\begin{equation}\label{eq:constr_N4}
% \text{Segment 1:}
\begin{cases}
    \B{r}_1(0)=\B{r}_i \\
    \dot{\B{r}}_1(0)=\B{v}_i\\
    \B{r}_1(T)=\B{r}_2(0) \\
    \dot{\B{r}}_1(T)=\dot{\B{r}}_2(0)
\end{cases}
% \text{Segment 2:}
\begin{cases}
    \B{r}_1(T)=\B{r}_2(0) \\
    \B{r}_2(T)=\B{r}_3(0) \\
    \dot{\B{r}}_1(T)=\dot{\B{r}}_2(0)\\
    \dot{\B{r}}_2(T)=\dot{\B{r}}_3(0)
\end{cases}
% \text{Segment 2:}
\begin{cases}
    \B{r}_2(T)=\B{r}_3(0) \\
    \B{r}_3(T)=\B{r}_4(0) \\
    \dot{\B{r}}_2(T)=\dot{\B{r}}_3(0)\\
    \dot{\B{r}}_3(T)=\dot{\B{r}}_4(0)
\end{cases}
% \text{Segment 3:}
\begin{cases}
    \B{r}_4(T)=\B{r}_f \\
    \dot{\B{r}}_4(T)=\B{v}_f\\
    \B{r}_3(T)=\B{r}_4(0) \\
    \dot{\B{r}}_3(T)=\dot{\B{r}}_4(0).
\end{cases}
\end{equation}
As some of them are repeated for the several segments, the integrated system is subject to following 10 constraints:
\begin{equation}\label{eq:constr_N4b}
\text{Combined constraints:}
\begin{cases}
    \B{r}_1(0)=\B{r}_i \quad \quad \quad \quad \quad \quad
    \dot{\B{r}}_1(0)=\B{v}_i \\
    \B{r}_1(T)=\B{r}_2(0)  \quad \quad \quad
    \dot{\B{r}}_1(T)=\dot{\B{r}}_2(0)\\
    \B{r}_2(T)=\B{r}_3(0) \quad \quad \quad
    \dot{\B{r}}_2(T)=\dot{\B{r}}_3(0)\\
    \B{r}_3(T)=\B{r}_4(0) \quad \quad \quad
    \dot{\B{r}}_3(T)=\dot{\B{r}}_4(0)\\
    \B{r}_4(T)=\B{r}_f \quad \quad \quad \quad \quad 
    \dot{\B{r}}_4(T)=\B{v}_f.
\end{cases}
\end{equation}
Thus, we need to distribute 10 polynomials among the support functions. We did this distribution according to
\begin{equation}\label{eq:sf4}
    \B{s}_1=\begin{Bmatrix}1,t,t^2\end{Bmatrix}\T; \quad
    \B{s}_2=\begin{Bmatrix}1,t\end{Bmatrix}\T; \quad
    \B{s}_3=\begin{Bmatrix}1,t\end{Bmatrix}\T; \quad \text{and} \quad
    \B{s}_4=\begin{Bmatrix}1,t,t^2\end{Bmatrix}\T.
% \text{Support functions:}
% \begin{cases}
%     \B{s}_1=\begin{Bmatrix}1,t,t^2\end{Bmatrix}\T\\
%     \B{s}_2=\begin{Bmatrix}1,t\end{Bmatrix}\T\\
%     \B{s}_3=\begin{Bmatrix}1,t\end{Bmatrix}\T\\
%     \B{s}_4=\begin{Bmatrix}1,t,t^2\end{Bmatrix}\T.
% \end{cases}
\end{equation}
The constrained functionals are then obtained following the same procedure outlined in the earlier sections: we use the support functions shown above (in Eq.~\eqref{eq:sf4}) into Eq.~\eqref{eq:generating_i0}, and then apply Eq.~\eqref{eq:generating_i0} into \eqref{eq:constr_N4b} to generate a linear system in $E_n$ for $n\in\llbracket 1:4 \rrbracket$. The \textit{constrained functional} can be then obtained by solving this linear system for $E_n$ and substituting the solution into Eq.~\eqref{eq:generating_i0}.

% \subsection{Segmentation and connection for \texorpdfstring{$N_s=5$}{Ns=5}}
\subsection{Segmenting into five parts}
\label{sec:segNs5}

In the case where $N_s=5$, the integrated system is subject to following 12 combined constraints representing the continuity constraints shown in Eq.~\eqref{eq:continuity} and the overdetermined constraints shown in Eq.~\eqref{eq:over_constr}:
\begin{equation}\label{eq:constr_N5b}
\text{Combined constraints:}
\begin{cases}
    \B{r}_1(0)=\B{r}_i \quad \quad \quad \quad \quad 
    \dot{\B{r}}_1(0)=\B{v}_i\\
    \B{r}_1(T)=\B{r}_2(0) \quad \quad \quad
    \dot{\B{r}}_1(T)=\dot{\B{r}}_2(0)\\
    \B{r}_2(T)=\B{r}_3(0) \quad \quad \quad
    \dot{\B{r}}_2(T)=\dot{\B{r}}_3(0)\\
    \B{r}_3(T)=\B{r}_4(0) \quad \quad \quad
    \dot{\B{r}}_3(T)=\dot{\B{r}}_4(0)\\
    \B{r}_4(T)=\B{r}_5(0) \quad \quad \quad
    \dot{\B{r}}_4(T)=\dot{\B{r}}_5(0)\\
    \B{r}_5(T)=\B{r}_f \quad \quad \quad \quad ~
    \dot{\B{r}}_5(T)=\B{v}_f.
\end{cases}
\end{equation}
We distribute 12 polynomials among the support functions according to
\begin{equation}\label{eq:sf5}
    \B{s}_1=\begin{Bmatrix}1,t,t^2\end{Bmatrix}\T; \quad
    \B{s}_2=\begin{Bmatrix}1,t\end{Bmatrix}\T; \quad
    \B{s}_3=\begin{Bmatrix}1,t\end{Bmatrix}\T; \quad
    \B{s}_4=\begin{Bmatrix}1,t\end{Bmatrix}\T; \quad \text{and} \quad
    \B{s}_5=\begin{Bmatrix}1,t,t^2\end{Bmatrix}\T.
% \begin{cases}
%     \B{s}_1=\begin{Bmatrix}1,t,t^2\end{Bmatrix}\T\\
%     \B{s}_2=\begin{Bmatrix}1,t\end{Bmatrix}\T\\
%     \B{s}_3=\begin{Bmatrix}1,t\end{Bmatrix}\T\\
%     \B{s}_4=\begin{Bmatrix}1,t\end{Bmatrix}\T\\
%     \B{s}_5=\begin{Bmatrix}1,t,t^2\end{Bmatrix}\T.
% \end{cases}
\end{equation}
The constrained functionals are then obtained following the same procedure outlined in earlier sections: we use the support functions shown above (in Eq.~\eqref{eq:sf5}) into Eq.~\eqref{eq:generating_i0}, and then apply Eq.~\eqref{eq:generating_i0} into \eqref{eq:constr_N5b} to generate a linear system in $E_n$ for $n\in\llbracket 1:5 \rrbracket$. The \textit{constrained functional} can be then obtained by solving this linear system for $E_n$ and substituting the solution into Eq.~\eqref{eq:generating_i0}.

% \begin{equation}\label{eq:constr_N5}
% % \text{Segment 1:}
% \begin{cases}
%     \B{r}_1(0)=\B{r}_i \\
%     \dot{\B{r}}_1(0)=\B{v}_i\\
%     \B{r}_1(T)=\B{r}_2(0) \\
%     \dot{\B{r}}_1(T)=\dot{\B{r}}_2(0)
% \end{cases}
% % \text{Segment 2:}
% \begin{cases}
%     \B{r}_1(T)=\B{r}_2(0) \\
%     \B{r}_2(T)=\B{r}_3(0) \\
%     \dot{\B{r}}_1(T)=\dot{\B{r}}_2(0)\\
%     \dot{\B{r}}_2(T)=\dot{\B{r}}_3(0)
% \end{cases}
% % \text{Segment 2:}
% \begin{cases}
%     \B{r}_2(T)=\B{r}_3(0) \\
%     \B{r}_3(T)=\B{r}_4(0) \\
%     \dot{\B{r}}_2(T)=\dot{\B{r}}_3(0)\\
%     \dot{\B{r}}_3(T)=\dot{\B{r}}_4(0)
% \end{cases}
% % \text{Segment 2:}
% \begin{cases}
%     \B{r}_3(T)=\B{r}_4(0) \\
%     \B{r}_4(T)=\B{r}_5(0) \\
%     \dot{\B{r}}_3(T)=\dot{\B{r}}_4(0)\\
%     \dot{\B{r}}_4(T)=\dot{\B{r}}_5(0)
% \end{cases}
% % \text{Segment 3:}
% \begin{cases}
%     \B{r}_5(T)=\B{r}_f \\
%     \dot{\B{r}}_5(T)=\B{v}_f\\
%     \B{r}_4(T)=\B{r}_5(0) \\
%     \dot{\B{r}}_4(T)=\dot{\B{r}}_5(0)
% \end{cases}
% \end{equation}

\subsection{The general case with many segments}
\label{sec:segNsn}

In the general case, we connect the segments through the continuity constraints (as shown in Eq.~\eqref{eq:continuity}) to connect the segments $n-1$ with $n$ and $n$ with $n+1$. The continuity constraints become
\begin{equation}\label{eq:seg_conn}
\begin{cases}
    \B{r}_{n-1}(T)=\B{r}_{n}(0) \\
    \B{r}_{n}(T)=\B{r}_{n+1}(0) \\
    \dot{\B{r}}_{n-1}(T)=\dot{\B{r}}_{n}(0)\\
    \dot{\B{r}}_{n}(T)=\dot{\B{r}}_{n+1}(0)
\end{cases}
\end{equation}

%Starting from $n=3$ and ending at the last even number for $n$ other than $N_s$, we derive the constrained functional from:
We derive next the constrained functionals for the case where $(N_s-2)/3$ is any positive integer ($N_s=5,8,11,14,...$). First, we derive the constrained functions associated to the internal segments, i.e. for all segments except the first and last one. 
This can be achieved using the following support functions:
\begin{equation}\label{eq:sfNsn}
\text{Support functions:}
\begin{cases}
    \B{s}_{n-1}=\{ \}\T\\
    \B{s}_{n}=\begin{Bmatrix}1,t,t^2,t^3\end{Bmatrix}\T\\
    \B{s}_{n+1}=\{ \}\T.
\end{cases}
\end{equation}
where $\{ \}\T$ means that no support function is adopted, thus the constrained functional is unaltered. In this case, the generating function shown in Eq. \eqref{eq:generating_i0} for the $n-1$, $n$, and $n+1$ segments become
\begin{equation}\label{eq:gen_cons_func}
\begin{cases}
    \B{r}_{n-1}(t)=\B{g}_{n-1}(t)\\
    \B{r}_{n}(t)=\B{g}_{n}(t) + E_n \, \B{s}_n \\%\eta_0 + \eta_1 t + \eta_2 t^2 + \eta_3 t^3 \\
    \B{r}_{n+1}(t)=\B{g}_{n+1}(t),
\end{cases}
\end{equation}
Equation \eqref{eq:gen_cons_func} is then combined with the segment connection shown in Eq.\eqref{eq:seg_conn} to derive the following constrained functionals:
\begin{equation}\label{eq:contr1}
\begin{cases}
\B{r}_{n-1}(t)=\B{g}_{n-1}(t)\\
    \B{r}_{n}(t)=\B{g}_{n}(t)+\B{g}_{n-1}(T)-\B{g}_{n}(0)+t \left(\dot{\B{g}}_{n-1}(T)-\dot{\B{g}}_{n}(0)\right)\\
    -\Big(2 T \dot{\B{g}}_{n-1}(T)+3 \B{g}_{n-1}(T)-2 T \dot{\B{g}}_{n}(0)-T \dot{\B{g}}_{n}(T)+3 \B{g}_{n}(T)-3 \B{g}_{n}(0)+T \dot{\B{g}}_{n+1}(0)-3 \B{g}_{n+1}(0)\Big)(\frac{t}{T})^2\\
    -\Big(-T \dot{\B{g}}_{n-1}(T)-2 \B{g}_{n-1}(T)+T \dot{\B{g}}_{n}(0)+T \dot{\B{g}}_{n}(T)-2 \B{g}_{n}(T)+2 \B{g}_{n}(0)-T \dot{\B{g}}_{n+1}(0)+2 \B{g}_{n+1}(0)\Big) (\frac{t}{T})^3\\
    \B{r}_{n+1}(t)=\B{g}_{n+1}(t)
\end{cases}
\end{equation}

It is important to note that this set of constrained functions is isolated from the first and last segments.
Equation \eqref{eq:contr1} is derived under the particular case where $(N_s-2)/3$ is a positive integer, and is valid from $n=3$ up to $n=N_s-2$. This means that the first and last constrained functionals are not included. 
The reason of deriving constrained functionals in the specific case where $(N_s-2)/3$ is a positive integer is that with this choice it is possible to make the second constrained functional independent of the third constrained functional and, analogously, the second last constrained functional is independent of the third last one. 
%it is possible to make the first constrained functional independent of the third constrained functional and, analogously, the last constrained functional is independent of the third last one. 
Since the second constrained functional is not function of the others with higher orders, it can be independently combined to derive the first constrained functional satisfying any type of constraint at the initial time of motion (for $t=0$) and the continuity constraints between the first and second segments shown in Eq.~\eqref{eq:seg_conn}. The last constrained functional can be obtained analogously.
Hence, the constrained functional for $\B{r}_1$ and $\B{r}_{N_s}$ depends on the constraints of the problem. For instance,
% in the particular Two-Point Boundary Value Problem (TPBVP) represented by the constraints
% \begin{equation}\label{eq:seg_conbvp}
% \begin{cases}
%     \B{r}(0)=\B{r}_i \\
%     \B{r}(T_f)=\B{r}_f,
% \end{cases}
% \end{equation}
% %$\B{r}(0)=\B{r}_i$ and $\B{r}(T_f)=\B{r}_f$,
% the first and last constrained functionals are derived as
% \begin{equation}\label{eq:general_tpbvp}
% \begin{cases}
%     \B{r}_{1}(t)=\B{g}_1(t)-\B{g}_1(0)+\B{r}_i-\left(-T \dot{\B{g}}_1(T)+2 \B{g}_1(T)-2 \B{g}_1(0)+T \dot{\B{g}}_2(0)-2 \B{g}_2(0)+2 \B{r}_i\right)t/T\\
% -\Big(T \dot{\B{g}}_1(T)-\B{g}_1(T)+\B{g}_1(0)-T \dot{\B{g}}_2(0)+\B{g}_2(0)-\B{r}_i\Big)(t/T)^2\\
%     \B{r}_{N_s}(t)=\B{g}_{N_s}(t)+\B{g}_{N_s-1}(T)-\B{g}_{N_s}(0)+ \Big(\dot{\B{g}}_{N_s-1}(T)-\dot{\B{g}}_{N_s}(0)\Big) t\\
% -\Big(T \dot{\B{g}}_{N_s-1}(T)+\B{g}_{N_s-1}(T)-T \dot{\B{g}}_{N_s}(0)+\B{g}_{N_s}(T)-\B{g}_{N_s}(0)-\B{r}_f\Big)(t/T)^2
% \end{cases}
% \end{equation}
% On the other hand, 
the constrained functionals representing $\B{r}_1$ and $\B{r}_{N_s}$ for the overdetermined constraints shown in Eq.~\eqref{eq:over_constr} are given by
\begin{equation}\label{eq:overd_case_Ns5}
\begin{cases}
    \B{r}_{1}(t)=\B{g}_1(t)-\B{g}_1(0)+\B{r}_i+t \left(\B{v}_i-\dot{\B{g}}_1(0)\right)+\frac{t^2 \left(2 T \dot{\B{g}}_1(0)+T \dot{\B{g}}_1(T)-3 \B{g}_1(T)+3 \B{g}_1(0)-T \dot{\B{g}}_2(0)+3 \B{g}_2(0)-2 T \B{v}_i-3 \B{r}_i\right)}{T^2}\\
    +\frac{t^3 \left(-T \dot{\B{g}}_1(0)-T \dot{\B{g}}_1(T)+2 \B{g}_1(T)-2 \B{g}_1(0)+T \dot{\B{g}}_2(0)-2 \B{g}_2(0)+T \B{v}_i+2 \B{r}_i\right)}{T^3}\\
    \B{r}_{N_s}(t)=\B{g}_4(T)+\B{g}_5(t)-\B{g}_5(0)+t \left(\dot{\B{g}}_2(T)-\dot{\B{g}}_5(0)\right)\\
    +\frac{t^2 \left(-2 T \dot{\B{g}}_2(T)-3 \B{g}_4(T)+2 T \dot{\B{g}}_5(0)+T \dot{\B{g}}_5(T)-3 \B{g}_5(T)+3 \B{g}_5(0)-T \B{v}_f+3 \B{r}_f\right)}{T^2}\\
    +\frac{t^3 \left(T \dot{\B{g}}_2(T)+2 \B{g}_4(T)-T \dot{\B{g}}_5(0)-T \dot{\B{g}}_5(T)+2 \B{g}_5(T)-2 \B{g}_5(0)+T \B{v}_f-2 \B{r}_f\right)}{T^3}.
\end{cases}
\end{equation}

The formulation presented above in this section has the advantage of showing a set of constrained functional in a general form independent of constraints applied at the initial and final times of motion. We adopted $N_s=5$ using the constrained functionals derived in this section and compared with the method for $N_s=5$ outlined in Sec. \ref{sec:segNs5}. Although the accuracies are similar, the numerical optimization convergence (of the procedure shown in Appendix) using the constrained functional shown in Sec. \ref{sec:segNs5} seems easier in comparison with using the constrained functionals derived in this section, i.e. it depends on a broader range of initial guess.
In fact, we encountered numerical convergence difficulties in finding solutions for Earth to Moon transfers using the constrained functionals derived in Sec. \ref{sec:segNsn}. These convergence problems arose for larger times of flight (more than 10 days) using large values of discretized time points ($N>200$), as will be explained later in Sec. \ref{sec:results}. On the other hand, we did not encounter these convergence issues using the constrained functional obtained in Sec. \ref{sec:segNs5}.
Possibly this happens because the connections (due to the continuity constraints) must be done using the formulation for the central segment ($\B{r}_n$) with respect to its neighbors ($\B{r}_{n-1}$ and $\B{r}_{n+1}$), whose constrained functionals remain unaltered. Hence, we do not recommend the general formulation shown in Eqs.~\eqref{eq:contr1} and \eqref{eq:overd_case_Ns5}. Instead, for $N_s>5$, we recommend distributing the support functions generalizing how they were distributed in Secs.  \ref{sec:segNs2}, \ref{sec:segNs3}, \ref{sec:segNs4}, and \ref{sec:segNs5}, which can be written as
\begin{equation}\label{eq:sfNs6}
\B{s}_n=
\begin{cases}
    % \B{s}_1=\begin{Bmatrix}1,t,t^2\end{Bmatrix}\T\\
    % \B{s}_n=\begin{Bmatrix}1,t\end{Bmatrix}\T \quad \text{for} ~ n\in\llbracket 2:N_s-1 \rrbracket\\
    % \B{s}_{N_s}=\begin{Bmatrix}1,t,t^2\end{Bmatrix}\T.
    \begin{Bmatrix}1,t,t^2\end{Bmatrix}\T \quad \text{if}\quad n=1,N_s\\
    \begin{Bmatrix}1,t\end{Bmatrix}\T \quad \quad ~ \text{if}\quad  n\ne 1,N_s
\end{cases}
\end{equation}
The constrained functionals can be derived using this distribution for the support functions. This can be done following the procedure outlined in the earlier sections. Although a general formulation could not be obtained in this case, this form might improve numerical convergence in comparison with the constrained functional shown in Eqs.~\eqref{eq:contr1} and \eqref{eq:overd_case_Ns5}. Thus, the numerical analysis shown in Sec. \ref{sec:results} are obtained using the constrained functional derived in Sec. \ref{sec:segNs5}, instead of those shown in Eqs.~\eqref{eq:contr1} and \eqref{eq:overd_case_Ns5} for the particular case where $N_s=5$.
% Note that we also adopted this distribution in Sec. \ref{sec:segNs2}, but since there is no central segment for the case $N_s=2$, we did not used the internal $\B{s}_n$ distribution shown in Eq.~\eqref{eq:sfNs6}.

\section{Application: Earth to Moon transfers}

In this section, we prepare to apply the methodology shown in this paper to evaluate transfers from the Earth to the Moon for times of flight between 1 and 15 days.
We show below in Sec. \ref{sec:tangential} how to take advantage of the overdetermined constraints to design Earth to Moon transfers, transforming nonlinear constraints under the linear vector approach shown in this paper.
These transfers are evaluated using the mathematical model given by the Circular Restricted Three-Body Problem (CRTBP) outlined in Sec. \ref{sec:crtbp}.
In order to prepare a numerical analysis of the methodology presented in this paper, we make a brief review on short-time transfers in Sec. \ref{sec:shorttime}.
%we apply the technique to evaluate time transfers for times of flight between 1 to 15 days.

\subsection{Tangential velocities and altitudes constraints}
\label{sec:tangential}

We now transform the initial and final positions and velocities into a combination of unknowns in order to guarantee that the initial position is located anywhere in a circle of radius $r_e$ around the Earth and the initial velocity is tangential to this initial circular orbit around Earth. An analogous procedure is done at the final time of motion such that the final position is located at a distance $r_m$ from the center of the Moon and the final velocity is tangential to the final circular orbit around the Moon. These assumptions can be represented 
%in the $x$-$y$ plane 
using rectangular coordinates by the transformation
\begin{equation}\label{eq:tang}
    \begin{cases}
        \B{r}_i=\{-d_e+r_e \cos \alpha, r_e \sin \alpha, 0\}\T\\
        \B{v}_i=\{-v_i \sin \alpha, v_i \cos \alpha, 0\}\T\\
        \B{r}_f=\{+d_m+r_m \cos \beta, r_m \sin \beta, 0\}\T\\
        \B{v}_f=\{-v_f \sin \beta, v_f \cos \beta, 0\}\T,
    \end{cases}
\end{equation}
where $\alpha$, $\beta$, $v_i$, and $v_f$ are four scalar unknowns to be numerically determined (see appendix for the optimization procedure).
Their physical meaning are explained next: $\alpha$ is the angle of the orbit around Earth measured with respect to the $x$ axis such that the first impulse is applied; $v_i$ represents plus or minus the magnitude of the initial velocity after the impulse; $\beta$ is the angle at the final orbit around the Moon to apply the final impulse; and $v_f$ is plus or minus the final velocity before the second impulse. They can be visualized in Fig. \ref{fig:graph1}.

The set of transformations proposed in Eq.~\eqref{eq:tang} takes advantage of the general vector formulation adopted in this paper and is very powerful to the transfer problem. Using the procedure explained above, we transform optimization results such as tangential velocities into constraints and use TFC to analytically embed them into the formulation of the problem. Tangential velocities here represent simultaneously a direction in the velocity and a position along an initial curve (initial transfer orbit) at $t=0$ and an analogous direction in the velocity and a position along a final curve (final transfer orbit) at $t=T$.
Thus, we eliminate the necessity of numerical optimization procedures to satisfy them, allowing a more efficient search in the space of solutions.
% Besides, every solution must departure from a circular orbit around the Earth at the initial time of motion and arrive at a circular 

\subsection{The Circular Restricted Three-Body Problem}
\label{sec:crtbp}

It is important to note that no equation of the dynamics is defined above. The methodology presented in this paper is general, and it can be applied to many systems subject to boundary constraints from different areas of research, such as physics, biology, engineering, etc. In this paper, we illustrate the methodology focusing on the particular space application of the orbit transfer problem. Although the presented technique can be applied to any mathematical model, like the bi-circular bi-planar four-body problem or a more accurate ephemeris based model, we chose the Circular Restricted Three-Body Problem (CRTBP) for simplicity reasons in the interpretation of the results. 

In the CRTBP, Earth and Moon rotates in circular motion around their barycenter with constant angular velocity $\B{\omega}$ perpendicular to the plane of motion. A rotating non-inertial frame of reference is defined such that Earth and Moon are fixed.
%along the $x$ axis, in rectangular coordinates. The $z$ axis is defined along their constant angular velocity $\B{\omega}$, and the $y$ axis completes an orthogonal system according to the right hand rule.
The equation of motion of a spacecraft for the $n^{\text{th}}$ segment is \cite{symon}
\begin{equation}
	\label{eq:crtbp}
	\frac{\text{d}^2\B{r}_n}{\text{d}t^2} = -2\B{\omega} \times \frac{\text{d}\B{r}_n}{\text{d}t} - \B{\omega} \times \left(\B{\omega} \times \B{r}_n \right) - \frac{\mu_e}{r_{n}^{'3}}  \B{r}'_{n} - \frac{\mu_m}{r_{n}^{''3}} \B{r}''_{n}  \quad \text{for} ~ n\in\llbracket 1:N_s \rrbracket,
\end{equation}
where $\B{r}_n$
%$\B{r}_n = \{x, y, z\}\T$
is the position vector of the spacecraft relative to the barycenter shown in Fig. \ref{fig:graph1}, $\B{r}'$ and $\B{r}''$ denote the position vectors of the spacecraft relative to the Earth and Moon, respectively, with norms $r'$ and $r''$. The gravitational parameters of the Earth and Moon are $\mu_e$ and $\mu_m$, respectively.
The values of the parameters
% \cite{simo1995book,YAGASAKI2004313,Yagasaki2004,topputo2013optimal} 
are shown in Table \ref{tab:parameters}, where $R$ is the distance between Earth and Moon.
\begin{table}[ht]
	\centering
	\begin{tabular}{lll}
		\hline
		$R$ & $3.84405000 \times 10^{8}$ & m \\[0.5ex]
		$R_s$ & $1.49460947424915\times 10^{11}$ & m \\[0.5ex]
		$\mu_e$ & $3.975837768911438\times10^{14}$ & m$^3$/s$^2$  \\[0.5ex]
		$\mu_m$ & $4.890329364450684\times10^{12}$ & m$^3$/s$^2$ \\[0.5ex]
		% $\mu_s$ & $1.3237395128595653 \times 10^{20}$ & m$^3$/s$^2$ \\[0.5ex] 
		$\omega$ & $2.66186135\times 10^{-6}$ & s$^{-1}$ \\[0.5ex]
		% $\omega_s$&$-2.462743433827215\times 10^{-6}$ & s$^{-1}$ \\[0.5ex]
		\hline
	\end{tabular}
    \caption{Values of the parameters, according to \cite{topputo2013optimal,simo1995book,YAGASAKI2004313}.}
	\label{tab:parameters}
\end{table}

\subsection{Short time transfer costs}
\label{sec:shorttime}

\begin{figure}
\centering\includegraphics[scale=0.45]{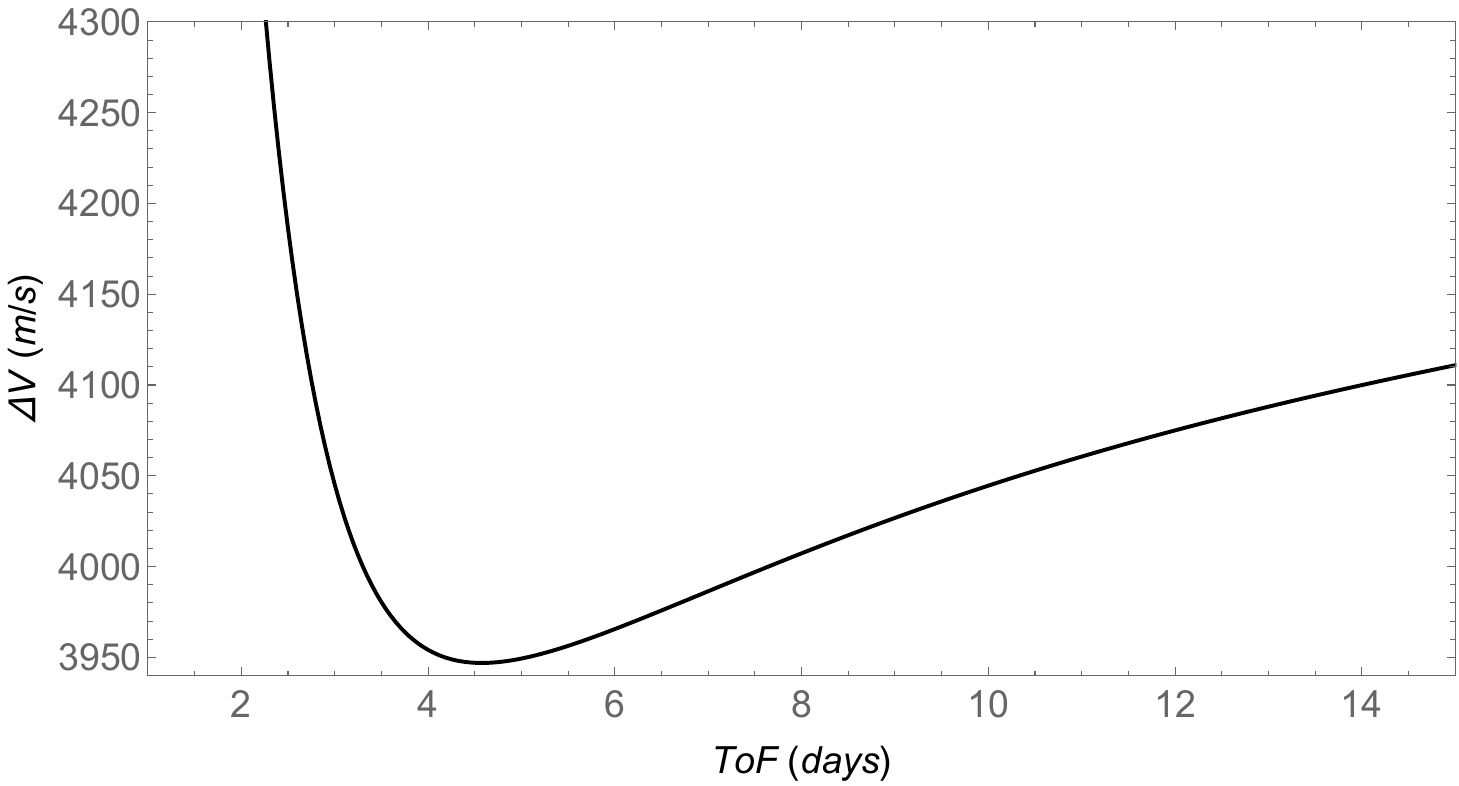}
\caption{Earth to Moon orbital transfer cost as a function of time of flight associated to the Pareto \textit{i} family of solutions. We adopt this family to compare segmented and non-segmented orbits.}
\label{fig:deltav}
\end{figure}

\begin{figure}
\centering\includegraphics[scale=0.45]{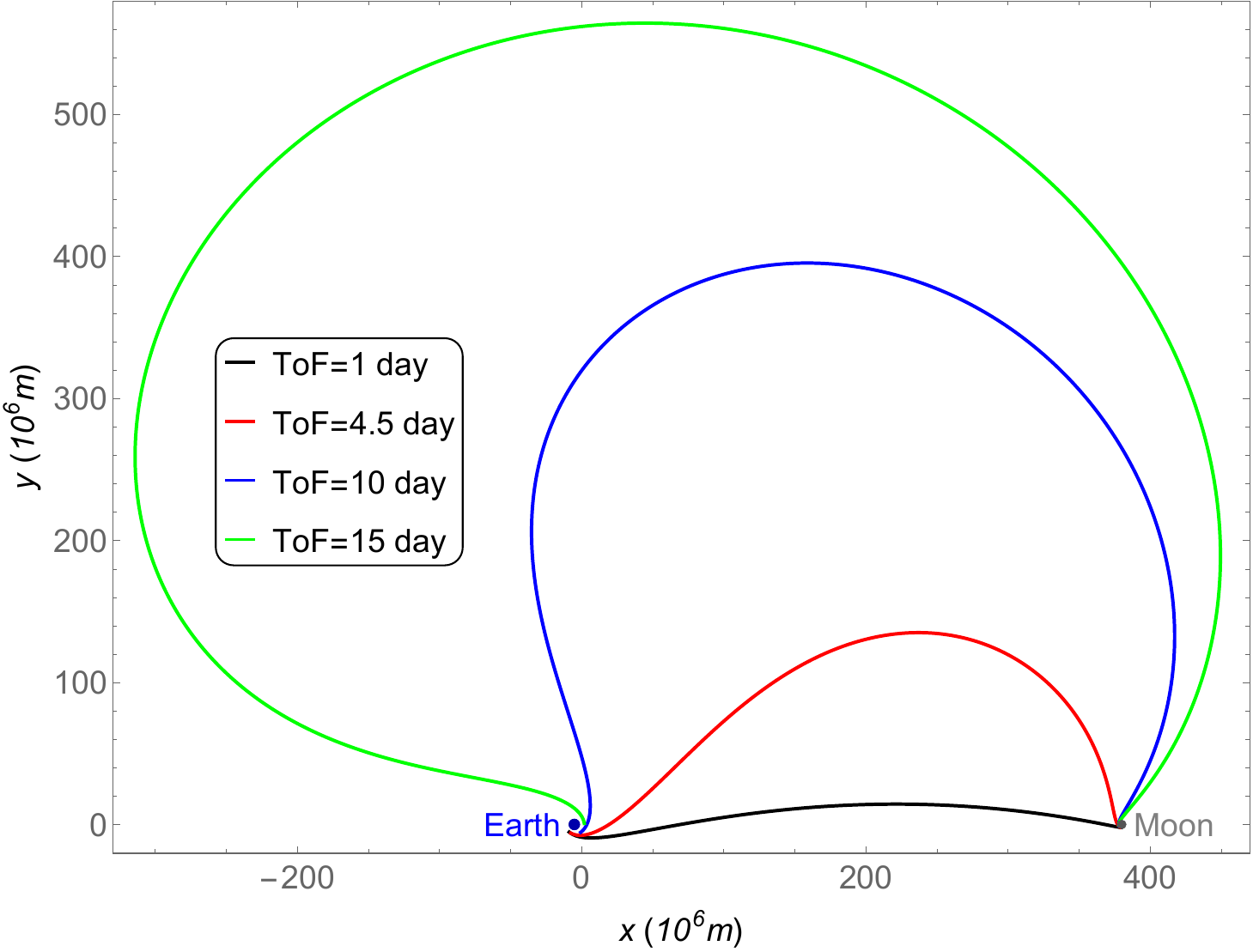}
\caption{Illustration of an Earth to Moon orbital transfer family for several values of times of flight shown in the rotating frame. The red transfer minimizes the costs.}
\label{fig:paretoiOrbits}
\end{figure}

We obtain the numerical results in Sec. \ref{sec:results} based on short-term transfers of a spacecraft from the Earth to the Moon, as briefly explained in Sec. \ref{sec:application}. The spacecraft is initially in a circular orbit around Earth of radius $r_e$ with altitude equal to 167 km, where the first impulse is applied. The spacecraft is then transferred to another circular orbit around Moon of radius $r_m$ with altitude equal to 100 km, where the second impulse is applied to circularize the orbit.
%These values are chosen in many works analyzing Earth-to-Moon transfers, e.g. \cite{topputo2013optimal}.
We adopt the Pareto optimal solution for short time transfers named as ``Pareto \textit{i}'' in \cite{topputo2013optimal}, whose minimum fuel equivalent delta-V is 3946.93 m/s for a time of flight (ToF) of 4.55 days \cite{fastTFC}. The orbits shown in Fig. \ref{fig:graph1} for several numbers of segmentation are associated to this minimum cost and time of flight. We use a continuation procedure to ``follow'' the family of solutions associated to this minimum for different times of flight in the range from 1 days to 15 days. The cost as a function of the ToF associated to this family is shown in Fig. \ref{fig:deltav} for illustrative purposes. Although another family arises close to ToF=15 days, named as Pareto \textit{ii} in \cite{topputo2013optimal}, we restricted our analyses to the first ``Pareto \textit{i}'' family of this reference for comparison purposes. Orbits for several values of ToF related to this family are shown in Fig. \ref{fig:paretoiOrbits} for illustrative purposes.

\section{Numerical Analysis}
\label{sec:results}

In this section, we analyze the results obtained from the application of the proposed methodology to design Earth-to-Moon transfers. We obtain the initial position and velocity using the segmented procedure shown in this paper to numerically solve for the unknowns (see Appendix) based on the overdetermined constraints explained in Sec. \ref{sec:oc}.
The parameter $N$ is associated to the number of discretization in the time domain to perform the optimization and obtain the unknowns through numerical convergence. The number of basis functions for the free function $m$ adopted in this work is $m=N$. Details on the these parameters are also shown in Appendix.

% The results for the TFC method were obtained using the Python programming language with the assistance of the TFC module \citep{tfc2021github} that uses automatic differentiation \citep{10.1145/355586.364791} and a just-in-time (JIT) compiler \citep{JaxGithub}.
% \tc{The Python package is chosen due to its high efficiency in solving the NLLS based in the \textit{jax.numpy} function \textit{linalg.pinv} (using singular-value decomposition), besides some automatized conveniences like transformation of the scale of time according to the chosen polynomial and construction of the solution based in the number of basis functions according to the type of array required by \textit{jax.numpy} in order to implement JIT. Elements like the type and number of basis functions are chosen by the user. The \textit{constrained expression} based in the chosen support functions and in the types and number of constraints must be obtained separately. The complete capabilities of the package can be seen in \citep{tfc2021github}.}

\subsection{Accuracy}

Since we specify the time of flight ToF, the procedure shown in this paper solves the system for the position, but also for the unknowns $\alpha$, $\beta$, $v_i$, and $v_f$. These unknowns are then 

We analyze the results based on the position error $\delta r$. It is defined as the difference between the initial position and velocity integrated for their respective ToF and the final position and velocity obtained through the segmented procedure to solve for the ``overdetermined boundary value problem'', as proposed above in this paper.

We obtain the position error $\delta r$ varying the ToF and the number of segmentation $N_s$, as shown in Table \ref{tab:data_N100} for the case where $N=100$. Position errors $\delta r$ larger than $10\%$ of Earth-Moon distance are suppressed from this Table. 
Although the technique based on TFC is accurate, the points become sparse while the time of flight increases. Thus, the order of magnitude of the position error $\delta r$ increases with the time of flight, as expected. 
%Note that a time domain transformation from $t$ to $\tau'$ is performed in the numerical optimization procedure to solve for the unknowns using Chebychev polynomials (see Appendix). The interval of $t$ is $\llbracket 0 : T \rrbracket$, while the time interval of $\tau'$ (and the Chebychev polynomials) is restricted to $\llbracket -1 : 1 \rrbracket$. On the other hand, the order of magnitude of the position error clearly decreases for higher number of segments.
%This time transformation makes a numerical contribution to decrease accuracy for higher values of the time $T$.
The non-segmented case represented by $N_s=1$ presents worst accuracy, and cannot be adopted for times of flight larger than 3 days. 
Astonishing three to four orders of magnitude for the position error is improved for bi-segmented orbits ($N_s=2$) in comparison with non-segmented ones (case $N_s=1$). In general, one order of magnitude is improved for three segments ($N_s=3$) in comparison with two segments ($N_s=2$). One order of magnitude is also improved for the case of four segments ($N_s=4$) in comparison with three segments ($N_s=3$). Finally, another order of magnitude is gained in the case of five segments ($N_s=5$) in comparison to the case of four segments ($N_s=4$). The analysis of the case where $N=200$ is done based on the data shown in Table \ref{tab:data_N200}. In this case, the position error improvements related to comparison between the segments are similar to the $N=100$ case shown in Table \ref{tab:data_N100}, although we can see about five orders of magnitude gain for the position error for $N=200$ in comparison to the case where $N=100$. It can also be noted that the order of magnitude of the position error cannot be lower than $10^{-7}$ meters and values close to this limit cannot be compared. This limitation seems to be due to machine error accuracy level. The position error $\delta r$ ins shown in Table \ref{tab:data_N400} in the case where there are 400 time discrete points ($N=400$) for the merged system, i.e. for every segment. About five order of magnitude is improved for the position error $\delta r$ for $N=400$ in comparison with $N=200$. Although the pattern of gain in accuracy in comparison among the number of segments $N_s$ seems to be somehow similar to the previous two tables, in this case the data are closer to machine level accuracy, thus the differences for comparisons among the number of segments are less sensible at this level, which occurs for lower times of flight.

\begin{table}[ht]
\centering
\caption{Earth-Moon orbital transfer position error $\delta r$ (in meters) for segmented (cases $N_s$ in $\llbracket 2:5\rrbracket$) and non-segmented (case $N_s=1$) orbits for 100 time discrete points ($N=100$). ToF is the time of flight of the transfer orbit.}
\label{tab:data_N100}
\begin{tabular}{cccccc}
\toprule
%ToF & $\delta r$ ($N_s=1$) & $\delta r$ ($N_s=2$) & $\delta r$ ($N_s=3$) & $\delta r$ ($N_s=4$) & $\delta r$ ($N_s=5$) \\
ToF &\multicolumn{5}{c}{$\delta r$ (m)}\\
\cline{2-6}
 & $N_s=1$ &  $N_s=2$ & $N_s=3$ & $N_s=4$ & $N_s=5$ \rule[-5pt]{0pt}{17pt}\\
\midrule
1.0 & $8.2975 \times 10^{2}$ & $1.7508 \times 10^{-1}$ & $3.3161 \times 10^{-3}$ & $2.0102 \times 10^{-4}$ & $1.3522 \times 10^{-5}$ \\
1.5 & $1.7590 \times 10^{4}$ & $4.3878 \times 10^{0}$ & $1.3246 \times 10^{-1}$ & $5.3116 \times 10^{-3}$ & $1.5145 \times 10^{-4}$ \\
2.0 & $4.9987 \times 10^{5}$ & $4.4500 \times 10^{1}$ & $3.8806 \times 10^{0}$ & $2.7692 \times 10^{-1}$ & $1.4800 \times 10^{-2}$ \\
2.5 & $3.7869 \times 10^{6}$ & $5.5364 \times 10^{2}$ & $2.4004 \times 10^{1}$ & $2.0493 \times 10^{0}$ & $5.0663 \times 10^{-1}$ \\
3.0 & $1.4910 \times 10^{7}$ & $1.4790 \times 10^{4}$ & $1.2793 \times 10^{2}$ & $3.4403 \times 10^{1}$ & $4.3381 \times 10^{-1}$ \\
3.5 & - & $9.1116 \times 10^{4}$ & $9.1811 \times 10^{1}$ & $3.4949 \times 10^{1}$ & $2.7593 \times 10^{1}$ \\
4.0 & - & $3.0295 \times 10^{5}$ & $5.1042 \times 10^{3}$ & $2.2250 \times 10^{2}$ & $7.5868 \times 10^{1}$ \\
4.5 & - & $7.3757 \times 10^{5}$ & $2.8593 \times 10^{4}$ & $3.7148 \times 10^{2}$ & $1.2628 \times 10^{1}$ \\
5.0 & - & $1.5244 \times 10^{6}$ & $8.8457 \times 10^{4}$ & $1.7042 \times 10^{3}$ & $2.5964 \times 10^{2}$ \\
5.5 & - & $2.8489 \times 10^{6}$ & $2.0084 \times 10^{5}$ & $1.0477 \times 10^{4}$ & $4.3656 \times 10^{2}$ \\
6.0 & - & $5.1090 \times 10^{6}$ & $3.8234 \times 10^{5}$ & $3.2613 \times 10^{4}$ & $5.2429 \times 10^{2}$ \\
6.5 & - & $9.1176 \times 10^{6}$ & $6.6617 \times 10^{5}$ & $7.5772 \times 10^{4}$ & $4.4952 \times 10^{3}$ \\
7.0 & - & $1.6034 \times 10^{7}$ & $1.1279 \times 10^{6}$ & $1.4796 \times 10^{5}$ & $1.4570 \times 10^{4}$ \\
7.5 & - & $2.7240 \times 10^{7}$ & $1.9226 \times 10^{6}$ & $2.5976 \times 10^{5}$ & $3.4854 \times 10^{4}$ \\
8.0 & - & - & $3.3318 \times 10^{6}$ & $4.3085 \times 10^{5}$ & $7.0524 \times 10^{4}$ \\
8.5 & - & - & $5.7946 \times 10^{6}$ & $7.0060 \times 10^{5}$ & $1.2829 \times 10^{5}$ \\
9.0 & - & - & $9.9008 \times 10^{6}$ & $1.1418 \times 10^{6}$ & $2.1800 \times 10^{5}$ \\
9.5 & - & - & $1.6365 \times 10^{7}$ & $1.8764 \times 10^{6}$ & $3.5542 \times 10^{5}$ \\
10.0 & - & - & $2.6021 \times 10^{7}$ & $3.0929 \times 10^{6}$ & $5.6636 \times 10^{5}$ \\
10.5 & - & - & - & $5.0606 \times 10^{6}$ & $8.9202 \times 10^{5}$ \\
11.0 & - & - & - & $8.1366 \times 10^{6}$ & $1.3962 \times 10^{6}$ \\
11.5 & - & - & - & $1.2765 \times 10^{7}$ & $2.1744 \times 10^{6}$ \\
12.0 & - & - & - & $1.9474 \times 10^{7}$ & $3.3643 \times 10^{6}$ \\
12.5 & - & - & - & $2.8882 \times 10^{7}$ & $5.1542 \times 10^{6}$ \\
13.0 & - & - & - & - & $7.7891 \times 10^{6}$ \\
13.5 & - & - & - & - & $1.1573 \times 10^{7}$ \\
14.0 & - & - & - & - & $1.6871 \times 10^{7}$ \\
14.5 & - & - & - & - & $2.4113 \times 10^{7}$ \\
15.0 & - & - & - & - & $3.3804 \times 10^{7}$ \\
\bottomrule
\end{tabular}
\end{table}

\begin{table}[ht]
\centering
\caption{Earth-Moon orbital transfer position error $\delta r$ (in meters) for segmented (cases $N_s$ in $\llbracket 2:5\rrbracket$) and non-segmented (case $N_s=1$) orbits for 200 time discrete points ($N=200$).}
\label{tab:data_N200}
\begin{tabular}{cccccc}
\toprule
ToF &\multicolumn{5}{c}{$\delta r$ (m)}\\
\cline{2-6}
 & $N_s=1$ &  $N_s=2$ & $N_s=3$ & $N_s=4$ & $N_s=5$ \rule[-5pt]{0pt}{17pt}\\
\midrule
1.0 & $1.2036 \times 10^{-3}$ & $7.3328 \times 10^{-7}$ & $1.9689 \times 10^{-6}$ & $6.8802 \times 10^{-7}$ & $9.1256 \times 10^{-7}$ \\
1.5 & $3.0264 \times 10^{-2}$ & $2.8973 \times 10^{-6}$ & $3.5317 \times 10^{-6}$ & $2.3427 \times 10^{-6}$ & $5.1236 \times 10^{-6}$ \\
2.0 & $1.0247 \times 10^{0}$ & $9.9606 \times 10^{-6}$ & $9.0003 \times 10^{-6}$ & $2.8750 \times 10^{-6}$ & $1.6964 \times 10^{-6}$ \\
2.5 & $4.3252 \times 10^{-1}$ & $6.0453 \times 10^{-4}$ & $2.7562 \times 10^{-5}$ & $5.3738 \times 10^{-6}$ & $1.8006 \times 10^{-5}$ \\
3.0 & $6.1231 \times 10^{1}$ & $1.9583 \times 10^{-2}$ & $5.0381 \times 10^{-5}$ & $1.4309 \times 10^{-5}$ & $3.3776 \times 10^{-5}$ \\
3.5 & $1.1605 \times 10^{2}$ & $2.4050 \times 10^{-2}$ & $1.1867 \times 10^{-3}$ & $3.7936 \times 10^{-5}$ & $2.7045 \times 10^{-5}$ \\
4.0 & $4.8711 \times 10^{2}$ & $7.7341 \times 10^{-1}$ & $1.8653 \times 10^{-3}$ & $1.3190 \times 10^{-5}$ & $2.6914 \times 10^{-6}$ \\
4.5 & $2.4498 \times 10^{3}$ & $1.2405 \times 10^{0}$ & $3.7498 \times 10^{-2}$ & $9.7259 \times 10^{-4}$ & $8.2856 \times 10^{-5}$ \\
5.0 & $1.2816 \times 10^{4}$ & $4.2463 \times 10^{0}$ & $4.6840 \times 10^{-2}$ & $2.2195 \times 10^{-3}$ & $1.3893 \times 10^{-4}$ \\
5.5 & $5.9955 \times 10^{4}$ & $2.0971 \times 10^{1}$ & $2.2836 \times 10^{-1}$ & $8.2127 \times 10^{-3}$ & $7.1016 \times 10^{-4}$ \\
6.0 & $2.2104 \times 10^{5}$ & $4.3009 \times 10^{1}$ & $9.4336 \times 10^{-1}$ & $4.0979 \times 10^{-2}$ & $2.3556 \times 10^{-3}$ \\
6.5 & $6.6354 \times 10^{5}$ & $4.9247 \times 10^{1}$ & $1.5260 \times 10^{0}$ & $6.0434 \times 10^{-2}$ & $5.1304 \times 10^{-4}$ \\
7.0 & $1.7084 \times 10^{6}$ & $1.4427 \times 10^{1}$ & $3.8041 \times 10^{-1}$ & $4.0321 \times 10^{-2}$ & $1.3476 \times 10^{-2}$ \\
7.5 & $3.7061 \times 10^{6}$ & $7.9887 \times 10^{1}$ & $4.6112 \times 10^{0}$ & $3.8221 \times 10^{-1}$ & $4.1480 \times 10^{-2}$ \\
8.0 & $6.4716 \times 10^{6}$ & $2.0341 \times 10^{2}$ & $1.4898 \times 10^{1}$ & $9.6406 \times 10^{-1}$ & $6.2412 \times 10^{-2}$ \\
8.5 & $6.8514 \times 10^{6}$ & $2.8705 \times 10^{2}$ & $2.9879 \times 10^{1}$ & $1.5043 \times 10^{0}$ & $2.2098 \times 10^{-2}$ \\
9.0 & $4.4546 \times 10^{6}$ & $2.8226 \times 10^{2}$ & $4.5893 \times 10^{1}$ & $1.3313 \times 10^{0}$ & $1.5659 \times 10^{-1}$ \\
9.5 & $7.4127 \times 10^{6}$ & $8.4498 \times 10^{2}$ & $5.6178 \times 10^{1}$ & $5.6574 \times 10^{-1}$ & $5.2904 \times 10^{-1}$ \\
10.0 & $1.7159 \times 10^{7}$ & $2.6805 \times 10^{3}$ & $5.2383 \times 10^{1}$ & $5.3566 \times 10^{0}$ & $1.0780 \times 10^{0}$ \\
10.5 & $3.2621 \times 10^{7}$ & $6.5077 \times 10^{3}$ & $2.9888 \times 10^{1}$ & $1.4081 \times 10^{1}$ & $1.6315 \times 10^{0}$ \\
11.0 & - & $1.3579 \times 10^{4}$ & $4.1769 \times 10^{1}$ & $2.7048 \times 10^{1}$ & $1.8060 \times 10^{0}$ \\
11.5 & - & $2.5704 \times 10^{4}$ & $1.1422 \times 10^{2}$ & $4.3596 \times 10^{1}$ & $9.6582 \times 10^{-1}$ \\
12.0 & - & $4.5367 \times 10^{4}$ & $1.9742 \times 10^{2}$ & $6.1464 \times 10^{1}$ & $1.7234 \times 10^{0}$ \\
12.5 & - & $7.5849 \times 10^{4}$ & $2.6519 \times 10^{2}$ & $7.6913 \times 10^{1}$ & $7.2049 \times 10^{0}$ \\
13.0 & - & $1.2136 \times 10^{5}$ & $3.8892 \times 10^{2}$ & $8.4926 \times 10^{1}$ & $1.6388 \times 10^{1}$ \\
13.5 & - & $1.8721 \times 10^{5}$ & $8.4977 \times 10^{2}$ & $8.0941 \times 10^{1}$ & $2.9818 \times 10^{1}$ \\
14.0 & - & $2.7998 \times 10^{5}$ & $1.9556 \times 10^{3}$ & $6.3987 \times 10^{1}$ & $4.7487 \times 10^{1}$ \\
14.5 & - & $4.0787 \times 10^{5}$ & $4.0574 \times 10^{3}$ & $5.1500 \times 10^{1}$ & $6.8528 \times 10^{1}$ \\
15.0 & - & $5.8127 \times 10^{5}$ & $7.6832 \times 10^{3}$ & $8.7221 \times 10^{1}$ & $9.1077 \times 10^{1}$ \\
\bottomrule
\end{tabular}
\end{table}

\begin{table}[ht]
\centering
\caption{Earth-Moon orbital transfer position error $\delta r$ (in meters) for segmented (cases $N_s$ in $\llbracket 2:5\rrbracket$) and non-segmented (case $N_s=1$) orbits for $N=400$.}
\label{tab:data_N400}
\begin{tabular}{cccccc}
\toprule
ToF &\multicolumn{5}{c}{$\delta r$ (m)}\\
\cline{2-6}
 & $N_s=1$ &  $N_s=2$ & $N_s=3$ & $N_s=4$ & $N_s=5$ \rule[-5pt]{0pt}{17pt}\\
\midrule
1.0 & $2.3461 \times 10^{-6}$ & $7.9538 \times 10^{-7}$ & $2.6704 \times 10^{-7}$ & $4.6472 \times 10^{-7}$ & $6.5308 \times 10^{-7}$ \\
1.5 & $4.5173 \times 10^{-6}$ & $5.7299 \times 10^{-6}$ & $2.2684 \times 10^{-6}$ & $5.1417 \times 10^{-6}$ & $5.3660 \times 10^{-6}$ \\
2.0 & $1.4935 \times 10^{-5}$ & $9.8358 \times 10^{-6}$ & $4.0573 \times 10^{-6}$ & $2.0329 \times 10^{-6}$ & $1.5082 \times 10^{-5}$ \\
2.5 & $1.7924 \times 10^{-5}$ & $1.0233 \times 10^{-5}$ & $9.8535 \times 10^{-6}$ & $9.6148 \times 10^{-7}$ & $9.8661 \times 10^{-6}$ \\
3.0 & $7.1295 \times 10^{-5}$ & $8.8132 \times 10^{-5}$ & $1.9638 \times 10^{-5}$ & $4.1977 \times 10^{-5}$ & $3.5554 \times 10^{-5}$ \\
3.5 & $6.3213 \times 10^{-5}$ & $4.8922 \times 10^{-5}$ & $1.0107 \times 10^{-4}$ & $5.9707 \times 10^{-5}$ & $3.1280 \times 10^{-5}$ \\
4.0 & $4.4348 \times 10^{-4}$ & $1.0317 \times 10^{-4}$ & $1.3785 \times 10^{-4}$ & $7.2568 \times 10^{-6}$ & $1.7419 \times 10^{-5}$ \\
4.5 & $3.1200 \times 10^{-3}$ & $6.8759 \times 10^{-5}$ & $2.7407 \times 10^{-5}$ & $7.8842 \times 10^{-6}$ & $6.3212 \times 10^{-7}$ \\
5.0 & $1.5363 \times 10^{-2}$ & $2.3360 \times 10^{-4}$ & $1.2836 \times 10^{-4}$ & $3.1132 \times 10^{-5}$ & $5.2728 \times 10^{-5}$ \\
5.5 & $6.8055 \times 10^{-3}$ & $1.3687 \times 10^{-4}$ & $7.7549 \times 10^{-5}$ & $1.3001 \times 10^{-4}$ & $3.3931 \times 10^{-5}$ \\
6.0 & $1.4966 \times 10^{-1}$ & $7.3275 \times 10^{-5}$ & $7.7953 \times 10^{-5}$ & $1.4033 \times 10^{-4}$ & $8.4002 \times 10^{-5}$ \\
6.5 & $3.7640 \times 10^{-1}$ & $7.1110 \times 10^{-5}$ & $5.4312 \times 10^{-5}$ & $6.7833 \times 10^{-6}$ & $1.2775 \times 10^{-4}$ \\
7.0 & $3.2627 \times 10^{-1}$ & $3.1544 \times 10^{-4}$ & $2.6233 \times 10^{-5}$ & $7.6898 \times 10^{-5}$ & $1.4836 \times 10^{-5}$ \\
7.5 & $6.5543 \times 10^{-1}$ & $5.7260 \times 10^{-5}$ & $1.6899 \times 10^{-4}$ & $7.5126 \times 10^{-5}$ & $2.8557 \times 10^{-5}$ \\
8.0 & $2.9977 \times 10^{0}$ & $2.0151 \times 10^{-5}$ & $1.5357 \times 10^{-4}$ & $7.9741 \times 10^{-5}$ & $1.4388 \times 10^{-4}$ \\
8.5 & $6.5837 \times 10^{0}$ & $2.1900 \times 10^{-5}$ & $6.1615 \times 10^{-5}$ & $5.6716 \times 10^{-5}$ & $4.3740\times 10^{-5}$ \\
9.0 & $9.7276 \times 10^{0}$ & $7.3229 \times 10^{-4}$ & $1.6550 \times 10^{-4}$ & $9.2390 \times 10^{-5}$ & $1.6194 \times 10^{-4}$ \\
9.5 & $8.9611 \times 10^{0}$ & $1.3254 \times 10^{-3}$ & $6.8423 \times 10^{-5}$ & $9.5378 \times 10^{-5}$ & $4.6948 \times 10^{-5}$ \\
10.0 & $1.9265 \times 10^{-1}$ & $1.2892 \times 10^{-3}$ & $2.4790 \times 10^{-5}$ & $2.4206 \times 10^{-4}$ & $3.4814 \times 10^{-5}$ \\
10.5 & $2.2932 \times 10^{1}$ & $1.5183 \times 10^{-3}$ & $7.7039 \times 10^{-5}$ & $1.4475 \times 10^{-5}$ & $4.7053 \times 10^{-6}$ \\
11.0 & $6.6110 \times 10^{1}$ & $7.9437 \times 10^{-3}$ & $1.7671 \times 10^{-4}$ & $1.9660 \times 10^{-4}$ & $1.6763 \times 10^{-4}$ \\
11.5 & $1.3460 \times 10^{2}$ & $1.9776 \times 10^{-2}$ & $2.1464 \times 10^{-4}$ & $4.9201 \times 10^{-5}$ & $1.1967 \times 10^{-4}$ \\
12.0 & $2.3829 \times 10^{2}$ & $3.5298 \times 10^{-2}$ & $2.8267 \times 10^{-5}$ & $1.3334 \times 10^{-4}$ & $6.5241 \times 10^{-5}$ \\
12.5 & $3.7332 \times 10^{2}$ & $4.6295 \times 10^{-2}$ & $4.6651 \times 10^{-5}$ & $5.4749 \times 10^{-5}$ & $2.6693 \times 10^{-4}$ \\
13.0 & $5.5041 \times 10^{2}$ & $4.0590 \times 10^{-2}$ & $5.5669 \times 10^{-4}$ & $1.5019 \times 10^{-4}$ & $5.3529 \times 10^{-5}$ \\
13.5 & $7.5450 \times 10^{2}$ & $4.9854 \times 10^{-3}$ & $7.1180 \times 10^{-4}$ & $1.5318 \times 10^{-4}$ & $1.1691 \times 10^{-4}$ \\
14.0 & $1.0657 \times 10^{3}$ & $1.0813 \times 10^{-1}$ & $1.6123 \times 10^{-3}$ & $1.7820 \times 10^{-5}$ & $3.5071 \times 10^{-4}$ \\
14.5 & $1.6954 \times 10^{3}$ & $3.0047 \times 10^{-1}$ & $2.2967 \times 10^{-3}$ & $2.2326 \times 10^{-5}$ & $1.2297 \times 10^{-4}$ \\
15.0 & $2.7686 \times 10^{3}$ & $5.9189 \times 10^{-1}$ & $1.1517 \times 10^{-3}$ & $6.2754 \times 10^{-6}$ & $3.6325 \times 10^{-4}$ \\
\bottomrule
\end{tabular}
\end{table}

\subsection{Estimating computational gains}

In this section we make an estimative analysis to compare the accuracy gains shown in Tables \ref{tab:data_N100}, \ref{tab:data_N200}, and \ref{tab:data_N400} with respect to their computational costs. The nonlinear least squares method required to numerically solve for the unknown variables depends on the evaluation (and the inversion operation) of the Jacobian. The number of elements of this Jacobian matrix is proportional to the number of discretized points $N$, but is also proportional to the number of unknowns $m$. In this paper, we adopt $m=N$, hence the number of elements of the Jacobian is proportional to $N^2$. The reader is invited to see appendix for details on these variables.

The computational costs involve many variables. When comparing two different codes, some of these variables can be controlled and measured, e.g. processor, memory, or computation time, but other are very difficult to be controlled and measured, such as the algorithm and the form of its translation to the language. For instance, comparing two different codes based on the computational time can be unfair, because it depends on how the different algorithms (different methods) are coded. For instance, one of them may not be fully optimized. In order to simplify comparison with several different algorithm and codes, in this paper, instead of using the computational time, we estimate the computational cost based on the number of elements of the Jacobian. This number represents then the costs associated to operations with this matrix, mainly to invert it.
The segmentation $N_s$ is proportional to the size (number of elements) of this matrix, because we stack the segments following the composition of the position shown in Eq.~\eqref{eq:r}. For instance, assuming $N$ fixed, if we adopt $N_s=2$, we double the elements of the Jacobian in comparison with the number of its elements in the non-segmented case where $N_s=1$. Thus, the number of elements is given by $N_s N^2$.
%On the other hand, we saw above that the number of elements of the Jacobian is proportional to $N^2$.

Based on the above explanations, we define the relative computational cost $C_c$ as the position error $\delta r$ multiplied by the number of elements of the Jacobian required to perform such operation as $C_c=\delta r N_s N^2$. The relative computational cost $C_c$ allows us to compensate for the increasing in computational costs for segmenting the orbit and to compare the results for different numbers of $N$, as can be seen in Table \ref{tab:data_costs1}.
Except in some particular cases (limited by machine level accuracy), this cost is drastically decreased for segmenting the orbit into two segments (case $N_s=2$) in comparison with non-segmented orbits (case $N_s=1$).
This cost is further decreased for larger numbers of segmentation, as can be seen in cases $N_s>2$.
Table \ref{tab:data_costs1} also shows that the relative computational cost $C_c$ tends to decrease with the increase of $N$, although this cost is proportional to $N^2$. This means that the gain in accuracy compensate for the loss in increasing the number of operations on the Jacobian.

%In this paper, the algorithms for different numbers of segments are differently coded, because the constrained functional becomes more complex for larger values of $N_s$. 
It is worth to point out that the cost $C_c$ does not consider the numerical implications of the forms of the constrained functionals, since they become more complex for larger values of $N_s$. For instance, the constrained functional shown in Eq.~\eqref{eq:Ns1} is simpler than the constrained functional shown in Eq.~\eqref{eq:cfN2}, which in turn is simpler than the constrained functional shown in Eq.~\eqref{eq:cfN3}, and so on. 

\begin{table}[ht]
\centering
\caption{Earth-Moon orbital transfer relative computational cost $C_c=\delta r N_s N^2$.}
\label{tab:data_costs1}
\renewcommand{\arraystretch}{0.75} %<- Adjusts the vertical spacing of the entire table
\begin{tabular}{ccccccc}
\toprule
%ToF & $\delta r$ ($N_s=1$) & $\delta r$ ($N_s=2$) & $\delta r$ ($N_s=3$) & $\delta r$ ($N_s=4$) & $\delta r$ ($N_s=5$) \\
ToF &$N$&\multicolumn{5}{c}{$C_c$ (m)}\\
\cline{3-7}
 && $N_s=1$ &  $N_s=2$ & $N_s=3$ & $N_s=4$ & $N_s=5$ \rule[-5pt]{0pt}{17pt}\\
\midrule
1.0 & 100 & $8.2975 \times 10^{6}$ & $3.5016 \times 10^{3}$ & $9.9482 \times 10^{1}$ & $8.0410 \times 10^{0}$ & $6.7612 \times 10^{-1}$ \\
1.0 & 200 & $4.8144 \times 10^{1}$ & $5.8662 \times 10^{-2}$ & $2.3627 \times 10^{-1}$ & $1.1008 \times 10^{-1}$ & $1.8251 \times 10^{-1}$ \\
1.0 & 400 & $3.7537 \times 10^{-1}$ & $2.5452 \times 10^{-1}$ & $1.2818 \times 10^{-1}$ & $2.9742 \times 10^{-1}$ & $5.2246 \times 10^{-1}$ \\
2.0 & 100 & $4.9987 \times 10^{9}$ & $8.9000 \times 10^{5}$ & $1.1642 \times 10^{5}$ & $1.1077 \times 10^{4}$ & $7.4002 \times 10^{2}$ \\
2.0 & 200 & $4.0988 \times 10^{4}$ & $7.9685 \times 10^{-1}$ & $1.0800 \times 10^{0}$ & $4.6000 \times 10^{-1}$ & $3.3928 \times 10^{-1}$ \\
2.0 & 400 & $2.3896 \times 10^{0}$ & $3.1475 \times 10^{0}$ & $1.9475 \times 10^{0}$ & $1.3012 \times 10^{0}$ & $1.2065 \times 10^{1}$ \\
3.0 & 100 & $1.4910 \times 10^{11}$ & $2.9580 \times 10^{8}$ & $3.8380 \times 10^{6}$ & $1.3761 \times 10^{6}$ & $2.1690 \times 10^{4}$ \\
3.0 & 200 & $2.4492 \times 10^{6}$ & $1.5666 \times 10^{3}$ & $6.0457 \times 10^{0}$ & $2.2895 \times 10^{0}$ & $6.7551 \times 10^{0}$ \\
3.0 & 400 & $1.1407 \times 10^{1}$ & $2.8202 \times 10^{1}$ & $9.4260 \times 10^{0}$ & $2.6865 \times 10^{1}$ & $2.8444 \times 10^{1}$ \\
4.0 & 100 & $1.1454 \times 10^{12}$ & $6.0590 \times 10^{9}$ & $1.5313 \times 10^{8}$ & $8.9002 \times 10^{6}$ & $3.7934 \times 10^{6}$ \\
4.0 & 200 & $1.9484 \times 10^{7}$ & $6.1873 \times 10^{4}$ & $2.2382 \times 10^{2}$ & $2.1105 \times 10^{0}$ & $5.3829 \times 10^{-1}$ \\
4.0 & 400 & $7.0957 \times 10^{1}$ & $3.3016 \times 10^{1}$ & $6.6171 \times 10^{1}$ & $4.6443 \times 10^{0}$ & $1.3935 \times 10^{1}$ \\
5.0 & 100 & $2.5649 \times 10^{12}$ & $3.0488 \times 10^{10}$ & $2.6537 \times 10^{9}$ & $6.8170 \times 10^{7}$ & $1.2982 \times 10^{7}$ \\
5.0 & 200 & $5.1265 \times 10^{8}$ & $3.3970 \times 10^{5}$ & $5.6208 \times 10^{3}$ & $3.5512 \times 10^{2}$ & $2.7785 \times 10^{1}$ \\
5.0 & 400 & $2.4580 \times 10^{3}$ & $7.4752 \times 10^{1}$ & $6.1614 \times 10^{1}$ & $1.9925 \times 10^{1}$ & $4.2186 \times 10^{1}$ \\
6.0 & 100 & $3.8070 \times 10^{12}$ & $1.0218 \times 10^{11}$ & $1.1470 \times 10^{10}$ & $1.3045 \times 10^{9}$ & $2.6215 \times 10^{7}$ \\
6.0 & 200 & $8.8417 \times 10^{9}$ & $3.4407 \times 10^{6}$ & $1.1320 \times 10^{5}$ & $6.5566 \times 10^{3}$ & $4.7113 \times 10^{2}$ \\
6.0 & 400 & $2.3946 \times 10^{4}$ & $2.3448 \times 10^{1}$ & $3.7417 \times 10^{1}$ & $8.9813 \times 10^{1}$ & $6.7202 \times 10^{1}$ \\
7.0 & 100 & $3.8721 \times 10^{12}$ & $3.2068 \times 10^{11}$ & $3.3838 \times 10^{10}$ & $5.9184 \times 10^{9}$ & $7.2850 \times 10^{8}$ \\
7.0 & 200 & $6.8336 \times 10^{10}$ & $1.1541 \times 10^{6}$ & $4.5649 \times 10^{4}$ & $6.4513 \times 10^{3}$ & $2.6951 \times 10^{3}$ \\
7.0 & 400 & $5.2203 \times 10^{4}$ & $1.0094 \times 10^{2}$ & $1.2592 \times 10^{1}$ & $4.9215 \times 10^{1}$ & $1.1869 \times 10^{1}$ \\
8.0 & 100 & - & $8.8907 \times 10^{11}$ & $9.9955 \times 10^{10}$ & $1.7234 \times 10^{10}$ & $3.5262 \times 10^{9}$ \\
8.0 & 200 & $2.5886 \times 10^{11}$ & $1.6273 \times 10^{7}$ & $1.7878 \times 10^{6}$ & $1.5425 \times 10^{5}$ & $1.2482 \times 10^{4}$ \\
8.0 & 400 & $4.7962 \times 10^{5}$ & $6.4483 \times 10^{0}$ & $7.3714 \times 10^{1}$ & $5.1037 \times 10^{1}$ & $1.1510 \times 10^{2}$ \\
9.0 & 100 & - & $2.1668 \times 10^{12}$ & $2.9702 \times 10^{11}$ & $4.5671 \times 10^{10}$ & $1.0900 \times 10^{10}$ \\
9.0 & 200 & $1.7819 \times 10^{11}$ & $2.2581 \times 10^{7}$ & $5.5072 \times 10^{6}$ & $2.1300 \times 10^{5}$ & $3.1318 \times 10^{4}$ \\
9.0 & 400 & $1.5564 \times 10^{6}$ & $2.3433 \times 10^{2}$ & $7.9440 \times 10^{1}$ & $5.9130 \times 10^{1}$ & $1.2956 \times 10^{2}$ \\
10.0 & 100 & - & $5.3306 \times 10^{12}$ & $7.8062 \times 10^{11}$ & $1.2372 \times 10^{11}$ & $2.8318 \times 10^{10}$ \\
10.0 & 200 & $6.8635 \times 10^{11}$ & $2.1444 \times 10^{8}$ & $6.2860 \times 10^{6}$ & $8.5706 \times 10^{5}$ & $2.1560 \times 10^{5}$ \\
10.0 & 400 & $3.0824 \times 10^{4}$ & $4.1253 \times 10^{2}$ & $1.1900 \times 10^{1}$ & $1.5492 \times 10^{2}$ & $2.7851 \times 10^{1}$ \\
11.0 & 100 & - & $3.9610 \times 10^{12}$ & $1.7727 \times 10^{12}$ & $3.2547 \times 10^{11}$ & $6.9809 \times 10^{10}$ \\
11.0 & 200 & $2.2598 \times 10^{12}$ & $1.0863 \times 10^{9}$ & $5.0123 \times 10^{6}$ & $4.3277 \times 10^{6}$ & $3.6120 \times 10^{5}$ \\
11.0 & 400 & $1.0578 \times 10^{7}$ & $2.5420 \times 10^{3}$ & $8.4822 \times 10^{1}$ & $1.2582 \times 10^{2}$ & $1.3411 \times 10^{2}$ \\
12.0 & 100 & - & $2.8426 \times 10^{12}$ & $3.6453 \times 10^{12}$ & $7.7897 \times 10^{11}$ & $1.6821 \times 10^{11}$ \\
12.0 & 200 & $5.9484 \times 10^{12}$ & $3.6294 \times 10^{9}$ & $2.3691 \times 10^{7}$ & $9.8342 \times 10^{6}$ & $3.4468 \times 10^{5}$ \\
12.0 & 400 & $3.8126 \times 10^{7}$ & $1.1295 \times 10^{4}$ & $1.3568 \times 10^{1}$ & $8.5338 \times 10^{1}$ & $5.2193 \times 10^{1}$ \\
13.0 & 100 & - & - & - & $1.6687 \times 10^{12}$ & $3.8946 \times 10^{11}$ \\
13.0 & 200 & - & $9.7092 \times 10^{9}$ & $4.6672 \times 10^{7}$ & $1.3588 \times 10^{7}$ & $3.2777 \times 10^{6}$ \\
13.0 & 400 & $8.8066 \times 10^{7}$ & $1.2989 \times 10^{4}$ & $2.6722 \times 10^{2}$ & $9.6123 \times 10^{1}$ & $4.2823 \times 10^{1}$ \\
14.0 & 100 & - & - & - & $3.2614 \times 10^{12}$ & $8.4356 \times 10^{11}$ \\
14.0 & 200 & - & $2.2399 \times 10^{10}$ & $2.3468 \times 10^{8}$ & $1.0238 \times 10^{7}$ & $9.4974 \times 10^{6}$ \\
14.0 & 400 & $1.7052 \times 10^{8}$ & $3.4602 \times 10^{4}$ & $7.7392 \times 10^{2}$ & $1.1405 \times 10^{1}$ & $2.8057 \times 10^{2}$ \\
15.0 & 100 & - & - & - & $6.0321 \times 10^{12}$ & $1.6902 \times 10^{12}$ \\
15.0 & 200 & - & $4.6501 \times 10^{10}$ & $9.2198 \times 10^{8}$ & $1.3955 \times 10^{7}$ & $1.8215 \times 10^{7}$ \\
15.0 & 400 & $4.4298 \times 10^{8}$ & $1.8940 \times 10^{5}$ & $5.5283 \times 10^{2}$ & $4.0163 \times 10^{0}$ & $2.9060 \times 10^{2}$\\
\bottomrule
\end{tabular}
\end{table}

\section*{Conclusions}

In this paper, we segmented orbits and showed how to connect them based on a combination of continuity and overdetermined constraints. The continuity constraints are linear and ensure smooth connections, while the overdetermined constraints - linear under the vector notation presented in this paper - facilitates numerical convergence in comparison with components constraints. The presented procedure allowed us to use TFC to analytically integrate these constraints into the orbital transfer problem.

Applications to Earth-Moon transfers show that solutions of the boundary value problem based on segmented orbits are several orders of magnitude more accurate in comparison with a similar technique based on non-segmented orbits. The increasing in the computational load is compensated with the analysis based on a relative computational cost index. Results showed several orders of magnitude gains in segmented the orbit in comparison with not segmented ones. Moreover, results showed that segmenting the orbit in more segments can further lower this cost. Thus, segmenting orbit under the \textit{overdetermined constraints} is highly recommended using the procedure shown in this paper due to a significant increasing in efficiency and robustness.

The methodology is written in this paper using a general formalism, where the technique can be applied to a broad range of boundary value problems.
% For instance, the overdetermined constraints can be extended to include a gravity-assist maneuver in a previous encounter with a massive body at a specified time.
In this sense, the presented technique can also be useful in several different areas of knowledge other than astrodynamics.

% \section*{Acknowledgements}
% I thank D. Mortari for sharing the TFC and T. Vaillant for discussions on continuity constraints.

\section*{Fundings}

This research was done with own resources.

% \bibliographystyle{alpha}
% \bibliographystyle{spmpsci}      % mathematics and physical sciences
% \bibliographystyle{spphys}       % APS-like style for physics
% \bibliography{elsarticle-num}   % name your BibTeX data base
\bibliographystyle{unsrt}
\bibliography{_references}

\section*{Appendix}

\subsection*{Numerical procedure}
\label{sec:numerical}

The constrained functional derived through Eq.~\eqref{eq:generating_i0}
%given in Eq.~(\ref{eq:cf2}), (\ref{eq:ce2c}), or (\ref{eq:cf3}), 
along with its first and second derivative are substituted into the differential equation shown in Eq.~(\ref{eq:crtbp}). This substitution results in the following unconstrained differential equation:
\begin{equation}\label{eq:unconst}
    \ddot{\B{g}}_n(t) - \B{a}_n'(\B{g}_1, \dot{\B{g}}_1, \ddot{\B{g}}_1,...,\B{g}_n, \dot{\B{g}}_n, \ddot{\B{g}}_n,...,\B{g}_{N_s}, \dot{\B{g}}_{N_s}, \ddot{\B{g}}_{N_s}, t) = \B{0}, \quad \text{for } n \in \llbracket 1 : N_s \rrbracket.
\end{equation}

The free function $\B{g}_n(t)$ is expressed as
\begin{equation}\label{eq:freefunction}
    \B{g}_n(t) = L_n \, \B{h}_n (\tau'(t)), \quad \text{for } n \in \llbracket 1 : N_s \rrbracket.
\end{equation}
where $L_n$ is a constant matrix of unknown coefficients of dimension $3 \times (m - k_n + 1)$, and $\B{h}(\tau')$ is a $(m - k_n + 1) \times 1$ vector composed of orthogonal basis functions, such as Chebyshev polynomials of the first kind \cite{abramowitzstegun1972}. Here, $m$ denotes the highest degree (i.e., truncation order) of the polynomial basis, and $k_n$ is the number of support functions adopted for the $n^{th}$ segment. The basis elements in $\B{h}_n$ are selected to be linearly independent of the support functions, resulting in a total of $(m - k_n + 1)$ terms.

Since Chebyshev polynomials are defined on the interval $-1 \leq \tau' \leq 1$, a change of the independent variable is required as
\[
\tau' = \frac{2t}{T} - 1,
\]
to map the time interval $[0, T]$ into the standard domain of the polynomials.

The time domain is discretized into $N + 1$ points using Chebyshev-Gauss-Lobatto nodes, as described in \cite{lanczos1988applied}:
\begin{equation}\label{eq:distr}
    t_j - t_0 = \left(1 - \cos\left(\frac{j \pi}{N}\right)\right) \frac{T}{2}, \quad \text{for } j \in \llbracket 0 : N \rrbracket.
\end{equation}

By combining the free function representation from Eq.~(\ref{eq:freefunction}) with the time discretization in Eq.~(\ref{eq:distr}), the differential equation in Eq.~(\ref{eq:unconst}) can be transformed into the discrete system of equations:
\begin{equation}\label{eq:eqmotion3}
    \B{a}_n''(L_1,...,L_{N_s},\alpha, \beta, v_i, v_f) = \B{0} \quad \text{for } n \in \llbracket 1 : N_s \rrbracket.
\end{equation}
where $\alpha$, $\beta$, $v_i$, and $v_f$ are the unknowns embedded in the constrained functionals through $\B{r}_i$, $\B{r}_f$, $\B{v}_i$, and $\B{v}_f$ according to Eq.~\eqref{eq:tang}.
%where $\B{a}_n'': \mathbb{R}^{3 \times (m - k_n + 1)} \to \mathbb{R}^{3 \times (N + 1)}$. %represents the residual function evaluated at the collocation nodes.
%Remember that we defined the position stacking its segments into columns in Eq.~\eqref{eq:r}.
%Although Eq.~\eqref{eq:crtbp} represents the motion in the $n^{th}$ segment, it can be analogously stacked to describe the equations of motion of the integrated system.
The integrated system is obtained by stacking Eq.~\eqref{eq:eqmotion3} for every value of $n$ in the interval $\llbracket 1 : N_s \rrbracket$. This procedure generates an integrated system where all the segments are merged as
\begin{equation}\label{eq:eqmotion4}
    \B{a}'''(L_1,...,L_{N_s},\alpha, \beta, v_i, v_f) = \B{0},
\end{equation}
where $\B{a}''': \mathbb{R}^{ 3 \times (N_s \times (m + 1) - k_{T})+4} \to \mathbb{R}^{N_s\times 3 \times (N + 1)}$ represents the residual function evaluated at the collocation nodes.
Equation \eqref{eq:eqmotion4} represents a system of $N_s\times3 \times (N + 1)$ nonlinear equations containing a number of unknowns given by $ 3 \times (N_s \times (m + 1) - k_{T})+4$, where $k_{T}=\sum_{n=1}^{N_s} k_n$ is the number of combined constraints. The solution of the resulting system of nonlinear equations - with the unknowns corresponding to the elements of $L_n$ for $n \in \llbracket 1 : N_s \rrbracket$ - is obtained using a nonlinear least squares optimization method \cite{NDE}. This method minimizes the norm of the residual matrix $\B{a}'''$. It can be noted that the total number of unknowns must be equal or lower than the number of equations, implying in $m\leq (3 k_T-4)/(3 N_s)+N$. In general, values of $m$ close to $N$ shows better numerical results. Thus, in this paper, we adopt $m=N$. We coded the methodology to obtain the numerical results using the Python language, built on automatic differentiation \cite{10.1145/355586.364791} with the aid of the TFC module \cite{tfc2021github}.

Once the optimal matrices $L_n$ for $n \in \llbracket 1 : N_s \rrbracket$ are found, the free function $\B{g}_n(t)$ can be reconstructed using Eq.~(\ref{eq:freefunction}). Substituting this free function back into the constrained functional yields the complete solution $\B{r}_n(t)$ of the $n^{th}$ segment, which can be composed using Eq.~\eqref{eq:r} to generate the complete orbit. It can be note that the optimal values of the unknowns $\alpha, \beta, v_i, v_f$ also completes the solution, because they represent the state at two points (at the initial and final times of motion). In fact, the variables $\alpha, \beta, v_i, v_f$ represent overdetermined solutions, because they come from overdetermined constraints.

After a convergence is found, we adopt a continuation method, explained next, to find solutions for a different time of flight $T$.
We first find values of $\textbf{L}_n ~ \text{for } n \in \llbracket 1 : N_s \rrbracket$ and $\alpha, \beta, v_i, v_f$ given by $\textbf{L}_n'$ and $\alpha', \beta', v_i', v_f'$, respectively, that are solutions to the transfer problem for the \textit{overdetermined constraints} in a time of flight $T$.
After, we use these values $\textbf{L}_n'$ and $\alpha', \beta', v_i', v_f'$ as initial guesses for the unknown coefficients matrices $\textbf{L}_n ~ \text{for } n \in \llbracket 1 : N_s \rrbracket$ and $\alpha, \beta, v_i, v_f$ to solve the transfer problem in a time of flight $T+\delta T$, where $\delta T$ is a displacement in time small enough to allow convergence to the next optimal solution of $\textbf{L}_n$ and $\alpha, \beta, v_i, v_f$. The value $\delta T=0.5$ days was adopted to obtain the numerical results shown in this paper.
%Now, we slightly vary the initial and final AEPs and/or the time of flight $T$, and use $\textbf{L}_1$ and $\B{\xi}_1$ as initial guess to 
%\textcolor{red}{After convergence is reached for a transfer between two points, these two points are slightly displaced towards the desired initial and final AEPs, and the solution to the transfer between the previous point is used as an initial guess to the next one. This process is iteratively repeated until convergence is reached for transfers between the desired initial and final AEPs. The variation in the time of flight $T$ is done similarly, by slightly varying $T$ and using the previous solution as an initial guess to the next one.}
By repeating this procedure, we then obtained the coefficients $\alpha, \beta, v_i, v_f$ for the desired transfer and time of flight, completing the solution.

\end{document}